\documentclass[a4paper,11pt, final]{article}
\usepackage{test,graphicx,subcaption,booktabs,paralist,multirow,framed}
\usepackage[bottom]{footmisc} % makes sure the footnote of first page is at the borrom
\usepackage[authoryear,round]{natbib}
\usepackage[usenames,dvipsnames]{xcolor}
\usepackage[colorlinks=true,citecolor=Mahogany,linkcolor=Mahogany,urlcolor=Mahogany]{hyperref}
\usepackage{doi}
\usepackage{mathtools}
\usepackage{bm}
\usepackage{ushort}
\usepackage{xspace} 
\usepackage{mathpazo} 
\usepackage[margin=1.in]{geometry}
\usepackage{setspace}
\usepackage{pifont} 
\usepackage{nicefrac} 
\usepackage{stmaryrd}
 
\spacing{1.25} 
\theoremstyle{plain}

\usepackage{float}
\definecolor{c1}{HTML}{4477AA}
\definecolor{c2}{HTML}{CC6677}

\usepackage{datetime}
\newdateformat{monthyeardate}{\monthname[\THEMONTH], \THEYEAR}
\usepackage{tabularx}

\newcolumntype{L}{>{\centering\arraybackslash}X}
\usepackage{siunitx}
\begingroup
\catcode`\_=11
\catcode`\:=11
 
\endgroup

\usepackage[obeyDraft]{todonotes}
\begin{document}
\title{\fontsize{17.2pt}{12pt} \textbf{The Trouble with Rational Expectations in Heterogeneous Agent Models: A Challenge for Macroeconomics}\footnote{This paper was originally delivered as the \emph{Economic Journal Lecture} at the Royal Economic Society's 2024 Annual Conference. I am grateful to the Royal Economic Society for this honor. I thank Klaus Adam, Marios Angeletos, Adrien Auclert, Isaac Baley, Adrien Bilal, Pedro Bordalo, Peter Bossaerts, Jean-Philippe Bouchaud, Tobias Broer, Andrew Caplin, Ben Enke, Xavier Gabaix, Laura Gati, Federico Gavazzoni, Joao Guerreiro, Ina Hajdini, Frank Heinemann, Cars Hommes, Zhen Huo, Alex Imas, Greg Kaplan, Alex Kohlhas, Per Krusell,  Søren Leth-Petersen, Chen Lian, Guido Lorenzoni, Charles Manski, Peter Maxted, Dmitry Mukhin, Emi Nakamura, Galo Nu\~{n}o, Ryan Oprea, Ricardo Reis, Matt Rognlie, Tom Sargent, Karthik Sastry, Andreas Schaab, Martin Schneider, Chengchun Shi, Alp Simsek, Johannes Stroebel, St\'{e}phane Surprenant, Chris Tonetti, Ben Van Roy, Mirko Wiederholt, Johannes Wohlfart, Moto Yogo, Jian-Qiao Zhu, and two anonymous referees for helpful comments. I am also grateful for excellent research assistance from Nicholas Tokay and support from the Leverhulme Trust.}}
\author{Benjamin Moll \thanks{London School of Economics}
}
\date{First version: November 2024	\\
	This version: \monthyeardate\today
	\\
\href{https://benjaminmoll.com/challenge/}{[latest version]}
}

	\maketitle

\begin{abstract}
%The thesis of this essay is that, in heterogeneous agent macroeconomics, the assumption of rational expectations about equilibrium prices is unrealistic, unnecessarily complicates computations, and should be replaced. This is because rational expectations imply that decision makers forecast equilibrium prices like interest rates by forecasting cross-sectional distributions. The result is an extreme version of the curse of dimensionality: dynamic programming problems in which the entire cross-sectional distribution is a state variable (``Master equation" a.k.a. ``Monster equation"). This is not only unrealistic but also limits the applicability of the heterogeneous-agent approach to some of the biggest questions in macroeconomics, namely those in which aggregate risk and non-linearities are key, like  financial crises. This troublesome feature of the rational expectations assumption poses a challenge: what should replace it? I outline three criteria that alternative approaches should satisfy:  (1) simplification of the computational solution, (2) consistency with empirical evidence, and (3) (some) immunity to the Lucas critique. I then discuss some potentially promising directions, including temporary equilibrium approaches, incorporating survey expectations, least-squares learning, and reinforcement learning.

The thesis of this essay is that, in heterogeneous agent macroeconomics, the assumption of rational expectations about equilibrium prices is unrealistic and should be replaced. Rational expectations imply that decision makers forecast equilibrium prices like interest rates by forecasting cross-sectional distributions. This leads to an extreme version of the curse of dimensionality: dynamic programming problems in which the entire distribution is a state variable (``Master equation" a.k.a. ``Monster equation"). Frontier computational methods struggle with these infinite-dimensional Bellman equations, making it implausible that real-world agents solve the associated decision problems. These difficulties also limit the applicability of the heterogeneous-agent approach to central questions in macroeconomics -- those involving aggregate risk and non-linearities such as  financial crises. This troublesome feature of the rational expectations assumption poses a challenge: what should replace it? I outline three criteria for alternative approaches:  (1) computational tractability, (2) consistency with empirical evidence, and (3) (some) immunity to the Lucas critique. I then discuss several promising directions, including temporary equilibrium approaches, incorporating survey expectations, least-squares learning, and reinforcement learning.
\end{abstract}

%\begin{enumerate}
%\item Add a paragraph to the introduction acknowledging the behavioral macro literature
%\item Stress that some of the proposed alternatives --especially those regarded successful to generate realistic nonlinearities and are based on complex theory-- may introduce their own computational
%challenges. The paper assumes that abandoning rational expectations will simplify computations,
%but in practice (or in principle at least), the alternative methods may introduce new
%layers of complexity that could be equally difficult to manage. Without explicit
%comparisons of computational costs, the claim that these approaches will be more
%"operational" remains speculative. Having speculative claims is absolutely fine, but the
%writing needs to reflect this fundamental uncertainty
%\end{enumerate}

\thispagestyle{empty}
\setcounter{page}{0}

\pagebreak

One of the key developments in macroeconomics research over the last three decades has been the incorporation of explicit heterogeneity into models of the macroeconomy. A benefit of this approach is the empirical discipline from matching macro models to micro data. But some of the biggest questions in macroeconomics remain out of reach of this approach. These are questions in which aggregate risk and aggregate non-linearities play a key role, for example the perennial question: why do developed economies experience infrequent but large boom-bust cycles like financial crises?\footnote{For example, \citet{he-krishnamurthy-restud,he-krishnamurthy-AER}, \citet{brunnermeier-sannikov}, \citet{bianchi}, \citet{mendoza}, \citet{elenev-landvoigt-vanN}, and  \citet{krishnamurthy-li}. Some recent work does take steps toward solving heterogeneous-agent versions of such models; see the related-literature discussion below.}

The root cause of these questions being out of reach is one particular assumption: that of rational expectations about equilibrium prices. Under rational expectations, decision makers in economies with heterogeneity forecast cross-sectional distributions in order to forecast prices. This results in Bellman equations in which a (typically) infinite-dimensional cross-sectional distribution is a state variable, an extreme version of the curse of dimensionality \citep{krusell-smith,denhaan}. This problem arises even though decision makers do not directly ``care about'' the distribution, i.e. it does not enter their objective functions; instead, as is standard in competitive equilibrium models, they only care about prices. The infinite-dimensional Bellman equation is called ``Master equation'' in the mathematics literature and has been aptly nicknamed the ``Monster equation'' due to its complexity.\footnote{To be clear, the Master equation is not specific to continuous-time models. Instead \emph{all} rational-expectations heterogeneous agent models with aggregate risk feature a Master equation. Indeed the Master equation was already there in \citet{krusell-smith} and \citet{denhaan}, though not fully spelled out and not with that name.} A recent literature has made impressive advances developing methods for solving this ``Monster equation"\footnote{See the end of this introduction as well as Section \ref{sec:HA_forecasting}.} but computations are still extremely costly, severely limiting the reach of heterogeneous-agent macroeconomics.\footnote{There is, of course, also the solution method of \citet{krusell-smith} and \citet{denhaan}, in which decision makers forecast prices by forecasting \emph{moments} of cross-sectional distributions. This approach has a bounded-rationality interpretation and simplifies computation relative to solving the full Master equation. I discuss similarities and differences with the approach advocated here further below.}

It is instructive to consider an example: suppose I lived in one of our models and wanted to forecast the evolution of future interest rates, say because I am considering taking out a mortgage to buy a house. According to our theories, I would realize that market-clearing interest rates depend on the entire cross-sectional distribution of different asset holdings in the economy (say the U.S. economy, the Euro area, or indeed the entire world economy). I would therefore forecast interest rates by forecasting this entire cross-sectional distribution.

This paper's main argument is that this forecasting behavior implied by rational expectations is not just computationally challenging -- it is conceptually implausible. If even our most advanced computational tools struggle with the ``Monster equation," how can we justify the assumption that real-world households and firms solve the associated decision problems?\footnote{For similar points, see the quotes by \citet{morgenstern}, \citet{manski}, and \citet{adam-marcet} below.} Macroeconomists are spending a lot of intellectual and computational horse power solving an unrealistically complex problem. Instead of solving ``Monster equations" we should replace the rational expectations assumption and solve the simpler equations corresponding to their actual price-forecasting behavior. In short, I argue against rational expectations on conceptual grounds (in addition to empirical ones): in complex environments with heterogeneous interacting agents, the assumption implies agent forecasting behavior that is extremely implausible.\footnote{A version of this argument arguably also applies in the representative-agent case. Indeed, similar criticisms have a long tradition in the economics literature -- see the literature discussion below. However, in my view, the argument is considerably stronger in the heterogeneous-agent case because of the model's complexity, in particular that the economy's state is an infinite-dimensional distribution.}

%Doing so would also open the door to applying the empirical and theoretical discipline of heterogeneous-agent macroeconomics to the study of economic booms and busts.

%This paper's main argument is that we should not make our lives so hard.\footnote{To be clear, what I mean is that the \emph{computational} complexity and cost should not so be so high. Better modeling expectations formation may well require more (and not less) mathematics. In Tom Sargent's words \citep{rolnick-sargent-interview}: \emph{``A rule of thumb is that the more dynamic, uncertain and ambiguous is the economic environment that you seek to model, the more you are going to have to roll up your sleeves, and learn and use some math. That’s life."}.}  

Departing from rational expectations would open the door to applying the empirical and theoretical discipline of heterogeneous-agent macroeconomics to the study of economic booms and busts. Of course, non-rational optimism and pessimism may also be what drives economic booms and busts in the first place \citep[][and related literature below]{minsky,kindleberger}. The idea that expectations may themselves be drivers of instability is therefore another argument for departing from rational expectations. Put another way, the promise of alternative approaches is to ``kill two birds with one stone'': to simplify the computation of heterogeneous-agent models with aggregate risk while, at the same time, making these models more realistic and more likely to generate interesting macroeconomic phenomena.

But what should replace rational expectations about equilibrium prices? A much more reasonable assumption is that decision makers forecast prices directly rather than indirectly by forecasting cross-sectional distributions. This holds the promise of sidestepping the curse of dimensionality. But how exactly do decision makers form these price expectations? I spell out three criteria for such alternatives to rational expectations  and discuss some promising directions for developing such alternative approaches.

Before developing my main thesis about rational expectations, I revisit the roots of this modeling device in the 1960s and 1970s through the lens of some key early writings like \citet{muth}, \citet{lucas-prescott} and \citet{lucas-neutrality}. Expectations about equilibrium prices played a (if not \emph{the}) key role in the development and popularization of the new hypothesis. These early writings also stressed that a key payoff of adopting rational expectations was to make models of the macroeconomy ``operational," meaning that these models could be readily computed, fit to data, and made clear predictions about the behavior of macroeconomic time series following policy changes -- see for example Lucas' FORTRAN quote below.

In heterogeneous-agent macroeconomics, the rational expectations assumption actually hinders rather than helps this ``operability" because it implies the extreme version of the curse of dimensionality discussed above. To explain the problem at a deeper, mathematical level, I use a specific example: the problem of forecasting wages and interest rates in an entirely standard textbook economy, a real business cycle model with household heterogeneity as in \citet{krusell-smith}. This model is a useful laboratory because the same logic applies to \emph{all} heterogeneous-agent models. The state of the economy is the cross-sectional joint distribution of income and wealth. Even though households and firms do \emph{not} directly ``care about'' this cross-sectional distribution, it becomes a state variable in their dynamic programming problems because they use it to forecast future wages and interest rates. As already noted, the Mean Field Games (MFG) literature has called this equation the  ``Master equation" \citep{cardaliaguet-delarue-lasry-lions} and this name and the associated formalism have also found their way back into economics in recent years \citep{macro-annual,schaab,bilal,gu-lauriere-merkel-payne}.

Next, I briefly review existing solution methods for heterogeneous-agent models with aggregate risk and explain why it may be worth to go back to the drawing board to develop alternative approaches. First, methods that employ ``MIT shocks" or linearize with respect to aggregate states are obviously not suitable for studying questions in which aggregate non-linearities are key. Second, methods that tackle the full rational expectations equilibrium (i.e. the Master equation) are what the criticism of this paper is aimed at: too much intellectual and computational horse power is aimed at solving an unrealistically complex problem. Third, methods that work with moments of the distribution \citep{krusell-smith,denhaan} lack realism for similar reasons as Master equation methods: also the idea that decision makers forecast means, variances, etc of distributions to forecast prices strains credibility. One variant of the Krusell-Smith approach is less subject to my criticism: the variant in which the ``Krusell-Smith moments" are the prices themselves. While this approach shares some similarities with the approach advocated below, I believe that we can do better, for example by incorporating key empirical evidence.

I then turn to the discussion of what should replace rational expectations about equilibrium prices. In place of a concrete alternative proposal,\footnote{I only know the problem, not the solution!} I spell out three criteria that such alternatives should satisfy. A common element in this class of alternative approaches is that decision makers form expectations about prices directly rather than indirectly via forecasting distributions. The key question is how to discipline the subjective probability distributions of equilibrium prices, i.e. how to navigate the ``wilderness of non-rational expectations" \citep{sargent-bounded,sargent-conquest,sargent-presidential}.\footnote{Sargent attributes the phrase to \citet{sims} but Sims actually writes about the ``wilderness of disequilibrium economics".} The goal of my three criteria is to help navigate this wilderness. The three criteria are: (1) computational tractability, (2) consistency with empirical evidence, and (3) endogeneity of beliefs to model reality (Lucas critique). An implication of Criterion 1 is that it eliminates from the list of candidate alternative approaches any non-rational expectations models requiring decision makers to compute the special case of rational expectations. Criteria 2 and 3 further narrow the set of admissible subjective beliefs. The alternative approach should also be compatible with global (rather than local) and recursive solution methods.

In the paper's final section, I outline some promising directions for developing alternative models of price beliefs that satisfy these three criteria. First, I discuss the fundamental concept of temporary equilibrium, i.e. competitive equilibrium given specified subjective beliefs, alongside the idea of internal rationality. Second, alternative models of price beliefs should incorporate the substantial body of empirical evidence on expectations formation, including work on survey expectations. I then turn to models of adaptive learning which are natural candidates for meeting the three criteria. I focus on two classes of learning algorithms: least-squares learning from the economics literature and reinforcement learning from the computer science literature. Reinforcement learning means learning value functions of incompletely-known Markov decision processes and has driven some impressive advances in artificial intelligence. The two forms of learning are linked because both are special cases of a broader class of stochastic approximation algorithms. Finally, I discuss the idea that decision makers may use heuristics and simplified, restricted forecasting models. Of course, implementing any of these approaches in practice will entail its own computational, empirical, and conceptual challenges, so my case for their promise remains necessarily speculative.

With this caveat, all approaches share a crucial feature: they should enable efficient \emph{global} solutions of heterogeneous-agent models with full cross-sectional distributions. Relaxing rational expectations simplifies agents’ decision problems \emph{inside} the model but does \emph{not} diminish the rich dynamics of the \emph{economy} itself, which still evolves stochastically and non-linearly, driven by the policy functions of forward-looking heterogeneous agents. This is precisely what holds the promise of generating the kind of rich macroeconomic dynamics -- such as booms and busts -- that rational-expectations heterogeneous-agent models have struggled to capture.

\noindent \paragraph{Related Literature.} Similar critiques of rational expectations have a long tradition in macroeconomics. In fact, they arguably predate the formal development of rational expectations: \citet{morgenstern} criticizes the assumption of perfect foresight -- the precursor to rational expectations -- for imposing that \emph{``the forward-looking individual must [...] know not only exactly the influence of his own actions on prices, but also that of all other individuals", ``must grasp all economic interrelations -- that is, must master economic theory"}, and for \emph{``the improbably high demands placed on the intellectual capacity of economic agents."} See  Appendix \ref{sec:morgenstern} for the original German passages.

Following this tradition, a huge literature in macroeconomics analyzes alternatives to the rational expectations paradigm, both empirically and theoretically. I will cover much of this work when I discuss promising directions in Section \ref{sec:directions}. For now, some useful survey articles and books are: \citet{sargent-bounded,sargent-conquest,sargent-presidential}, \citet{woodford-AR}, \citet{angeletos-huo-sastry}, \citet{bordalo-gennaioli-shleifer-JEP}, \citet{enke-cognitive-turn}, \citet{ortoleva}, and \citet{baley-veldkamp} on the theoretical side and \citet{armantier-etal}, \citet{manski-macroannual}, \citet{weber-dacunto-gorodnichenko-coibion}, \citet{dacunto-weber}, and \citet{fofana-patzelt-reis} on the empirical side. Also see the recent Handbook of Economic Expectations \citep{handbook-expectations} that covers both theory and empirics.

As already noted, I will also review in detail various existing numerical solution methods for heterogeneous-agent models (see Section \ref{sec:HA_forecasting}). Closest to the approach advocated here is the bounded-rationality approach of \citet{krusell-smith} and \citet{denhaan}, specifically the variant in which the ``Krusell-Smith moments" are the prices themselves.

One main motivation of this essay is the pragmatic desire to compute heterogeneous-agent models with aggregate risk and non-linearities so as to study phenomena like financial crises. While largely out of reach for the heterogeneous-agent literature, a few recent contributions have taken steps in this direction \citep[e.g.][]{FV-hurtado-nuno,gopalakrishna-gu-payne}. Yet high computational costs mean that theories with aggregate non-linearities remain highly stylized and cannot fully speak to key empirical evidence. Rather than solving ``Monster equations", we should solve simpler, more realistic ones. Rather than ``taming the curse of dimensionality" \citep{FV-nuno-perla}, we should sidestep it. 

Finally, as already noted, non-rational expectations may also be the drivers of instability generating booms and busts in the first place. This idea has a long tradition going back to the verbal discussions by \citet{minsky} and \citet{kindleberger}. Recent examples of theories in which departures from rational expectations generate boom-bust-cycles include \citet{branch_evans_2011},  \citet{adam-marcet-beutel}, \citet{barberis-handbook}, \citet{williams-escape,williams-instability}, \citet{bouchaud-morelli-etal},  \citet{bordalo-gennaioli-shleifer-terry}, \citet{maxted-sentiment}, and \citet{krishnamurthy-li}. Agent-based modeling (ABM) is another modeling approach that aims to generate interesting macroeconomic phenomena of this type.\footnote{See, for example, the surveys by \citet{bouchaud-crises} and \citet{axtell-farmer}, the latest volume of the Handbook of Computational Economics \citep{handbook-CE}, or \citet{geanakoplos-etal-ABM}, \citet{bouchaud-gualdi-etal}, and \citet{muellbauer-etal-data-driven}.} Relative to heterogeneous-agent macroeconomics, ABM typically departs from all of (1) rational expectations, (2) optimization, (3) market clearing. I instead advocate for departing from (1) with regard to equilibrium prices but retaining (2) and (3).

\noindent \paragraph{Roadmap.}
Section \ref{sec:roots} puts the thesis of this essay into perspective by revisiting some key early writings on rational expectations. Section \ref{sec:HA} explains why, in heterogeneous-agent models, the assumption of rational expectations about equilibrium prices is unrealistic and should be replaced. Section \ref{sec:replace} poses the challenge ``what should replace rational expectations?" and spells out three criteria that alternative approaches should satisfy. Section \ref{sec:directions} discusses some promising directions for developing alternative models of price beliefs. Section \ref{sec:conclusion} concludes.

\section{Back to the roots of rational expectations: it was all about equilibrium prices \label{sec:roots}}

This section revisits the roots of the rational expectations modeling device in the 1960s and 1970s through the lens of some writings that were key to its development and popularization like \citet{muth}, \citet{lucas-prescott} and \citet{lucas-neutrality}. Expectations about equilibrium prices played a (if not \emph{the}) key role in these developments. In contrast, the recent theoretical expectations literature in macroeconomics mostly focuses on modeling expectations about exogenous variables. The early writings also stressed that a key payoff of adopting rational expectations was to make models ``operational" with an emphasis on computation.

\subsection{\citet{muth} ``Rational Expectations and the Theory of Price Movements" \label{sec:muth}}
Rational expectations were first proposed by John Muth in this 1961 paper. Already the paper's title, \emph{``Rational Expectations and the Theory of Price Movements"}, indicates that price expectations play a key role. To \emph{``best explain what the hypothesis is all about"}, Muth first uses a particular example: expectations about equilibrium prices in a single market. He starts with the following simple demand-supply system (see his Equation 3.1 in Section 3):\footnote{To make the equations easier to read, I denote demand by $D_t$ instead of $C_t$ and supply by $S_t$ instead of $P_t$. Both demand and supply are expressed as deviations from their long-run means, hence the potentially negative values.}
\begin{equation}\label{eq:muth}
\begin{split}
D_t &= - \beta p_t \qquad \quad \  (\mbox{Demand}),\\
S_t &= \gamma p_t^e + u_t \qquad (\mbox{Supply}),\\
D_t &= S_t \qquad \qquad \ \ \ (\mbox{Market equilibrium}).
\end{split}
\end{equation}
The first equation is the time-$t$ demand curve with demand $D_t$ a decreasing function of the price $p_t$. The second equation is the supply curve with supply $S_t$ an increasing function of the \emph{expected} price $p_t^e$ and a stochastic, potentially serially correlated supply shock $u_t$. The third equation is the market clearing condition which equates demand and supply. Production takes place with a one-period lag, i.e. suppliers make their production decision at time $t-1$, and hence $p_t^e$ is the expectation of $p_t$ based on information available at time $t-1$.

Muth's rational expectations hypothesis is a particular assumption about the formation of the price expectation $p_t^e$. Muth verbally states his hypothesis as follows: \emph{``Expectations of firms (or, more generally, the subjective probability distribution of outcomes) tend to be distributed, for the same information set, about the prediction of the theory (or the `objective' probability distributions of outcomes)."} Applying this idea to the demand-supply system \eqref{eq:muth}, rational expectations is the assumption that the subjective price expectation equals the objective, model-generated price expectation
$$p_t^e = \mathbb{E}_{t-1}[p_t],$$
where the expectations operator $\mathbb{E}_{t-1}$ integrates over the objective, model-generated price distribution. Rational expectations are ``model-consistent" in this sense and this may be a more appropriate terminology \citep{simon-rationality}. In his linear model, Muth solves for the rational price expectation as follows. First, the equilibrium price of the demand-supply system \eqref{eq:muth} is given by the reduced form
\begin{equation}\label{eq:reduced_form}
p_t  = -\frac{\gamma}{\beta} \mathbb{E}_{t-1}[p_t] - \frac{1}{\beta}u_t
\end{equation}
so that the equilibrium price is low if firms' price expectation is high (because this means that supply is high). Second, taking the time $t-1$ expectation $\mathbb{E}_{t-1}$ of this equilibrium price yields $\mathbb{E}_{t-1}[p_t]  = -\frac{\gamma}{\beta} \mathbb{E}_{t-1}[p_t] - \frac{1}{\beta}\mathbb{E}_{t-1}[u_t]$ or
$$p_t^e = \mathbb{E}_{t-1}[p_t]= - \frac{1}{\beta+\gamma} \mathbb{E}_{t-1}[u_t]$$
which is Muth's equation (3.5). Hence the expected equilibrium price depends only on the expected fundamental, the expectation of the supply shock $\mathbb{E}_{t-1}[u_t]$. A particularly simple case is when $u_t$ is serially uncorrelated and furthermore $\mathbb{E}_{t-1}[u_t]=0$. In this case the rational price expectation simply equals zero, $p_t^e = \mathbb{E}_{t-1}[p_t]=0$. From \eqref{eq:reduced_form} the actual equilibrium price equals $p_t=-u_t/\beta$ which is different from zero and moves around according to the realization of the supply shock $u_t$. But firms correctly understand that these supply shocks average out. Analogous results obtain when $u_t$ is serially correlated.\footnote{\label{foot:rho}Assuming $u_{t} = \rho u_{t-1} + \varepsilon_{t}$ for an i.i.d. mean-zero random variable $\varepsilon_t$ and correlation coefficient $\rho$, the rational price expectation is $\mathbb{E}_{t-1}[p_t]= - \frac{1}{\beta+\gamma} \rho u_{t-1}$ and the actual equilibrium price is $p_t  = -\frac{1}{\beta+\gamma}\rho u_{t-1} - \frac{1}{\beta}\varepsilon_t$ from \eqref{eq:reduced_form}.}

Two appealing features of this approach are as follows. First, to make predictions about equilibrium prices and quantities, no measurement of price expectations is needed because these are instead implied by the theory. Second, because firms understand the determination of equilibrium prices (i.e. they ``understand the structure of the economy") and equilibrium prices depend only on the fundamental $u_t$ (and model parameters), modeling firms' price expectations requires only the stochastic process of the fundamental $u_t$.

\subsection{In Bob Lucas' words \label{sec:bob}}
Dissatisfaction with prior modeling of price expectations, often via backward-looking ``adaptive expectations," was a key impetus for the popularization of the rational expectations hypothesis in the 1960s and 1970s. In his Nobel Prize Lecture, \citet{lucas-nobel} describes the situation:

\bigskip 
\emph{``[1960s-style macroeconometric models] implied behavior of actual equilibrium prices and incomes that bore no relation to, and were in general grossly inconsistent with, the price expectations that the theory imputed to individual agents. [...] This modeling inconsistency became more and more glaring. John Muth’s (1961) ``Rational Expectations and the Theory of Price Movements" focused on this inconsistency, and showed how it could be
removed by taking into account the influences of prices, including future prices, on  quantities and simultaneously the effects of quantities on equilibrium prices."}
\bigskip 

Similarly, Lucas' 1980 methodological review ``Methods and Problems in Business Cycle Theory" \citep{lucas-methods} stresses the desire to reconcile subjective and objective probability distributions of equilibrium prices:

\bigskip 
\emph{``One needs a principle to reconcile the price distributions implied by the market equilibrium with the distributions used by agents to form their own views of the future.  John Muth noted that [...] these distributions could not differ in a systematic way. His term for this latter hypothesis was rational expectations."}

\subsection{\citet{lucas-prescott} ``Investment under Uncertainty" \label{sec:lucas-prescott}}
Lucas and Prescott's important 1971 paper extended rational expectations to dynamic equilibrium models, showed how to formulate a recursive rational expectations equilibrium in a stochastic market economy using dynamic programming techniques, and thus laid the foundation for much of modern macroeconomics up to this day.\footnote{See \citet{sargent-palgrave} for a useful discussion of \citet{lucas-prescott}.}

The paper models the equilibrium of a single industry in which a large number of competitive firms make production and investment decisions in the face of random demand shocks. Specifically, there is an exogenously-given downward-sloping demand curve for the industry's output with a demand shifter that follows a Markov process which leads to fluctuations in the output price faced by firms.

The key difficulty then is how to model expectations about future demand and thus equilibrium prices. Lucas and Prescott write:

\bigskip
\emph{``From the viewpoint of firms in this industry, forecasting future demand means simply forecasting future output prices. The  usual way to formulate this problem is to postulate some forecasting rule for firms,  which in turn generates some pattern of investment behavior, which in turn, in conjunction with industry demand, generates an actual price series.}

\noindent \emph{Typically the forecasting rule postulated takes the form of anticipated prices being a fixed function of past prices -- `adaptive expectations.' But it is clear that [this implies that]  price forecasts and actual prices will have different probability distributions, and this difference will be persistent, costly to forecasters, and readily correctible.}

\noindent \emph{To avoid this difficulty, we shall, in this paper, go to the opposite extreme, assuming that the actual and anticipated prices have the same probability distribution, or that price expectations are rational."}
\bigskip

This last sentence is followed by a footnote stating: \emph{``This term is taken from \citet{muth}, who applied it to the case where the expected and actual price (both random variables) have a common mean value. Since Muth's discussion of this concept applies  equally well to our assumption of a common distribution for these random variables, it seems natural to adopt the term here."} While Muth had already verbally defined rational expectations in terms of a common distribution (see Section \ref{sec:muth}), he had only applied his concept to linear models in which only means matter, and so \citet{lucas-prescott} first applied rational expectations to entire probability distributions, i.e. in the more general way in which it is commonly used today. Perhaps even more importantly, Lucas and Prescott first showed how to formulate a recursive rational expectations equilibrium in a stochastic market economy using dynamic programming techniques -- their key equation (9) is a Bellman equation which indirectly defines the recursive competitive equilibrium by maximizing social surplus.

%$$v(k,u) = \sup_{x \geq 0} \left\{s(k,u) - x + \beta \int v \left[k h\left(\frac{x}{k},z \right) \right]p(dz,u) \right\} $$

In summary, also from Lucas and Prescott's pathbreaking paper, it becomes clear that the development of rational expectations was all about improving the modeling of expectations about equilibrium prices.

\subsection{\citet{lucas-neutrality} ``Expectations and the Neutrality of Money" \label{sec:lucas-neutrality}}
While the economics of Lucas' neutrality paper are, of course, different from the investment paper discussed in the preceding section, the methodology -- specifically the construction of a recursive rational expectations equilibrium -- is very similar. Lucas is after a Phillips curve, the relation between the change in the price level (inflation) and real output, in an overlapping-generation economy with two physically separated markets (islands). The key variable in his model economy is the equilibrium price level $p$ and the key difficulty is how young individuals form expectations about the future equilibrium price level $p'$ when deciding how much to consume, work, and save for future consumption. This difficulty is exactly analogous to the difficulty of firms predicting future output prices in \citet{lucas-prescott}. So is the solution which \citet{lucas-neutrality} explains as follows:

\bigskip
\emph{``Equilibrium prices and quantities will be characterized mathematically as functions defined on the space of possible states of the economy, which are in turn characterized as finite dimensional vectors. This characterization permits a treatment of the relation of information to expectations which is in some ways much more satisfactory than is possible with conventional adaptive expectations hypotheses."}
\bigskip

In a bit more detail, Lucas first writes down the young's optimization problem in terms of an unspecified (subjective) probability distribution $F$ for the future price level $p'$ -- see his equation (3.7). He then notes: \emph{``The state of the economy is fully described by the three variables $(m,x,\theta)$ [the money supply, the money growth rate, and the fraction of young people on the first island]. [...] If this is so, one can express the equilibrium price as a function $p(m,x,\theta)$ on the space of possible states."}\footnote{Also see the useful discussion of Lucas' paper by \citet{chari-architect} who writes \emph{``The key to the technical contribution is that prices are thought of as \emph{functions} of the state of the economy, where the state is the stock of money and the distribution of young people across islands."}} A recursive rational expectations equilibrium is then a price \emph{function} $p(m,x,\theta)$ that satisfies the optimality condition of the young's optimization problem (3.7) but with the expectation taken with respect to the joint distribution $G$ of future $(m',x',\theta')$ conditional on the current price $p(m,x,\theta)$ in place of the price-level distribution $F$. Lucas writes: \emph{``We have dispensed with unspecified distribution $F$, taking the expectation instead with respect to the well-defined distribution $G$."}

In a paper written for a conference held at the Federal Reserve Bank of Minneapolis to celebrate the 25th anniversary of the publication of \citet{lucas-neutrality}, \citet{sargent-lucas-nonneutrality} again emphasizes the importance of rational expectations as a device for modeling expectations about endogenous equilibrium variables: \emph{``The victory of rational expectations owes to its beauty and its utility: the economy with which it eliminates what we had thought were free variables -- peoples' expectations about endogenous variables -- while adding \emph{no} free parameters, but bringing instead cross-equation and cross-frequency restrictions."}

\subsection{The Payoff: ``Operational" Macro Theories \label{sec:operational}}
The early writings reviewed in this section stress one particular payoff of adopting the rational expectations device: that they make models of the macroeconomy ``operational." By this they mean that models with rational expectations can be readily computed, fit to data, and make clear predictions about policy counterfactuals.

\citet{lucas-methods} writes: \emph{``Our task as I see it [...] is to write a FORTRAN program that will accept specific economic policy rules as `input' and will generate as `output' statistics describing the operating characteristics of time series we care about, which are predicted to result from these policies."} Similarly, directly after the passage cited in section \ref{sec:lucas-prescott}, \citet{lucas-prescott} state: \emph{``[By imposing rational expectations,] we obtain an operational investment theory linking current investment to observable current and past explanatory variables, rather than to ``expected" future variables which must, in practice, be replaced by various ``proxy variables."}

This ``operability" of rational expectations models, including that they could be computed relatively easily, was a key catalyst of their fast diffusion in macroeconomics. I will return to this section's themes -- expectations about equilibrium prices and operability -- later on in the paper.

Finally, while modern macroeconomics owes much to the rational expectations hypothesis, its importance should not be overstated. Other aspects like micro foundations and general equilibrium are arguably more important for today's macroeconomics and the label ``rational expectations revolution" may be unhelpful. See \citet{werning-miracles} for a good discussion.

\section{The trouble with rational expectations about equilibrium prices in heterogeneous agent models \label{sec:HA}}

With this historical background, I now turn to the main thesis of this essay: in heterogeneous agent macroeconomics, the assumption of rational expectations about equilibrium prices is unrealistic and should be replaced. This is because rational expectations implies that decision makers (unrealistically) forecast equilibrium prices like interest rates by forecasting cross-sectional distributions, producing an extreme version of the curse of dimensionality.

It is instructive to consider the following thought experiment: suppose I lived in one of our models and wanted to forecast the evolution of future interest rates, say because I am considering taking out a mortgage to buy a house. According to our theories, I would realize that market-clearing interest rates depend on the entire cross-sectional distribution of different asset holdings in the economy (say the U.S. economy, the Euro area, or indeed the entire world economy). I would therefore forecast interest rates by forecasting this entire cross-sectional distribution. It seems self-evident that real-world individuals do not do this.

To understand the problem at a deeper, mathematical level, I will use a specific example: the problem of forecasting wages and interest rates in a textbook heterogeneous-agent economy. As a warm-up, I will start with a representative-agent economy (the real business cycle model). I will then add heterogeneity as in the economy studied by \citet{krusell-smith}.

\subsection{Forecasting equilibrium prices in representative agent models}
Time is discrete and indexed by $t=0,...,T$ where $T$ may be finite or infinite.\footnote{\label{foot:finiteT}The reason for choosing to nest finite $T$ is that the special case of only two time periods $t=0,1$ is useful below.} The only source of uncertainty is aggregate productivity $z_t$ which follows a Markov process. A representative consumer has preferences over consumption $c_t$ and hours worked $n_t$ given by
$$\mathbb{E}_0\sum_{t=0}^T \beta^t U(c_t,n_t),$$
where $U$ is a standard utility function and $0<\beta<1$ is a discount factor. A representative firm uses capital $k_t$ and labor $\ell_t$ to produce output according to a technology subject to productivity shocks
$$z_t F(k_t,\ell_t).$$
Capital accumulates according to $k_{t+1} = i_t + (1-\delta)k_t$ where $i_t$ is investment and $0<\delta<1$ the depreciation rate. The economy's resource constraints are:
$$c_t + i_t = z_t F(k_t,\ell_t), \qquad \ell_t = n_t, \quad \mbox{all} \ t=0,...,T.$$
This completes the description of this simple economy, a real business cycle (RBC) model. The only not-entirely-standard feature is that the time horizon $T$ may be finite (see footnote \ref{foot:finiteT}).

The definition of a competitive equilibrium is also standard. To fix ideas, I consider the particular decentralization in which firms own the economy's capital stock which they finance by issuing risk-free bonds to households who use these bonds as their saving vehicle. A competitive equilibrium is then quantities and prices $\{w_t,r_t\}$ such that:
\begin{enumerate}
\item Households maximize taking as given $\{w_t,r_t\}$
\begin{equation}\label{eq:household_RA}
\max_{\{c_t,n_t,a_{t+1}\}} \mathbb{E}_0\sum_{t=0}^T \beta^t U(c_t,n_t) \quad \mbox{s.t.} \quad c_t + a_{t+1} = w_t n_t + (1+ r_t) a_t.
\end{equation}
\item Firms maximize taking as given $\{w_t,r_t\}$
\begin{equation}\label{eq:firms}
\max_{\{x_t,\ell_t,k_{t+1}\}} \mathbb{E}_0\sum_{t=0}^T R_{0 \to t}^{-1}\left(z_t F(k_t,\ell_t) - w_t \ell_t - i_t  \right) \quad \mbox{s.t.} \quad k_{t+1} = i_t + (1-\delta)k_t,
\end{equation}
with $R_{0 \to t} = \prod_{s=1}^t (1+r_s)$.
\item Markets clear
\begin{equation}\label{eq:resource_RA}
k_t = a_t, \qquad \ell_t = n_t, \quad \mbox{all} \ t=0,...,T.
\end{equation}
\end{enumerate}
The key difficulty, which is the focus of this paper, is that optimizing households and firms need to forecast future wages and interest rates $\{w_t,r_t\}$. To flesh this out, let us focus on wages for now (the logic for interest rates will be exactly analogous). Household optimization \eqref{eq:household_RA} yields labor supply $n_t = n(w_t,a_t)$ that depends on the wage and the household's state variables, here wealth $a_t$. Analogously, firm optimization  \eqref{eq:firms}, yields labor demand  
$\ell_t = \ell(w_t,k_t,z_t)$. From market clearing, therefore, the equilibrium wage is a function of the economy's states, aggregate capital $k_t$ and aggregate productivity $z_t$:
$$w_t = w^*(k_t,z_t).$$
How then do households and firms forecast future wages? Given rational expectations, they do so by forecasting the economy's state variables $(k_t,z_t)$. This is because they understand the structure of the economy and, in particular, the mapping from $(k_t,z_t)$ to equilibrium wages $w_t$. An analogous logic holds for forecasting equilibrium interest rates $r_t = r^*(k_t,z_t)$. This use of ``equilibrium price functions" is exactly like in \citet{lucas-neutrality} -- see Section \ref{sec:lucas-neutrality}. It is worth noting that, even in this extremely stylized representative-agent economy, a rational expectations equilibrium is the solution to a relatively complicated fixed-point problem and that the informational requirements for decision makers are substantial and arguably unrealistic.

\paragraph{Solution methods for the representative agent case.} For questions where aggregate risk and non-linearities matter, one needs a global solution method. There are two main types.\footnote{Many well-developed \emph{local} solution methods exist for representative-agent models with aggregate risk, see e.g. the popular Dynare toolbox.}  The first tackles the competitive equilibrium directly. Even in a  simple RBC model like the one I just outlined, global solution methods for the competitive equilibrium are actually surprisingly challenging to implement. A classic method uses dynamic programming with the ``Big $K$, little $k$" trick \citep[e.g.][]{prescott-mehra,lucas-models,RMT} but there are various other ones; see, for example, \citet{maliar-maliar} and \citet{cao-luo-nie}.

The second, more common, approach is to invoke the second welfare theorem and solve for the equilibrium allocation via a social planner’s problem. This approach works only in special cases, but the key payoff is that the planner’s problem does not feature prices, so it sidesteps the main difficulty emphasized above -- expectations about equilibrium prices.

\subsection{Forecasting equilibrium prices in heterogeneous agent models \label{sec:HA_forecasting}}
Next consider the same economy but with household heterogeneity. Specifically, as in \citet{aiyagari} and \citet{krusell-smith}, assume that households are ex-ante identical but subject to idiosyncratic income risk in the form of labor productivity $y_{it}$ which follows a Markov process. This results in ex-post heterogeneity in household wealth and productivity $(a_{it},y_{it})$ and I denote the cross-sectional joint distribution of income and wealth by $G_t(a,y)$.

Analogous to the definition in the representative-agent economy, a competitive equilibrium is quantities and prices $\{w_t,r_t\}$ such that
\begin{enumerate}
\item Households maximize taking as given $\{w_t,r_t\}$
\begin{equation}\label{eq:household_HA}
\max_{\{c_{it},n_{it},a_{it+1}\}} \mathbb{E}_0\sum_{t=0}^T \beta^t U(c_{it},n_{it}) \quad \mbox{s.t.} \quad c_{it} + a_{it+1} = w_t y_{it} n_{it} + (1+ r_t) a_{it}.
\end{equation}
\item Firms maximize taking as given $\{w_t,r_t\}$ -- the same problem as in the representative-agent economy \eqref{eq:firms} since the economy's production side is unchanged.
\item Markets clear
\begin{equation}\label{eq:resource_HA}
k_t = \int a dG_t(a,y), \qquad \ell_t = \int n_t(a,y) y dG_t(a,y), \quad \mbox{all} \ t=0,...,T.
\end{equation}
\end{enumerate}
Importantly, in this competitive equilibrium, households and firms do \emph{not} ``care about" the cross-sectional distribution $G_t$, i.e. it does not enter their objective functions. Instead they only care about \emph{prices} $\{w_t,r_t\}$.

As before, the key difficulty is that households and firms need to forecast these future prices. Focusing again on wages, household optimization \eqref{eq:household_HA} now yields \emph{household-specific} labor supplies $n_{it} = n(w_t,a_{it},y_{it})$.\footnote{Labor supplies are household-specific except in the knife-edge case without income effects (e.g. GHH utility).} Labor demand is identical to the representative-agent case and given by $\ell_t = \ell(w_t,k_t,z_t)$. From market clearing, therefore, the equilibrium wage is a function of the economy's states, now the \emph{entire cross-sectional distribution $G_t$} in addition to aggregate productivity $z_t$:
\begin{equation}\label{eq:wage_function}
w_t = w^*(G_t(a,y),z_t).
\end{equation}
The same is true for the equilibrium interest $r_t = r^*(G_t(a,y),z_t)$. This follows from the optimality condition of the representative firm $r_t = z_t F_k(k_t,\ell_t) - \delta$ and that $\ell_t = \int n_t(a,y) y dG_t(a,y)$.

\paragraph{A generic feature of heterogeneous-agent models.} While this dependence of equilibrium prices on the entire cross-sectional distribution $G_t$ can be avoided in special cases,\footnote{
%For example, in the model above, the equilibrium interest rate $r_t$ depends only on the aggregate capital stock, $r_t = r^*(k_t,z_t)$ with $k_t = \int a dG_t(a,y)$. This follows from the market clearing condition for capital in \eqref{eq:resource_HA} and the optimality condition of the representative firm. Similarly
For example, if labor supply were perfectly inelastic \citep[as in the benchmark economy of][]{krusell-smith} or if preferences ruled out income effects, the equilibrium wage and interest rate would only depend on the aggregate capital stock $k_t$. However, even in this case, it would still be true that $G$ becomes a state variable in household's decision problem. \citet{rios-rull} explains it nicely (substituting my notation for his): \emph{``The pair $(z,k)$ is not, in general, a sufficient statistic for $k'$: tomorrow's capital $k'$ depends on the whole distribution of wealth $G$. Depending on how wealth is distributed, aggregate capital will be different tomorrow, except where individual decision rules are linear in $a$, which is not the usual case."}} it is important to note that this distributional dependence is a \emph{generic} feature of heterogeneous-agent models. To this end, denote individual state variables by a vector $x_{it}$ with distribution $G_t(x)$, exogenously evolving aggregate state variables by a vector $z_t$, and prices by a vector $p_{t}$ so that the model we considered above is the special case in which $x_{it}=(a_{it},y_{it})$ and $p_t=(w_t,r_t)$. Then, generically, equilibrium prices satisfy 
\begin{equation}\label{eq:price_generic}
p_t = \mathcal{P}^*(G_t(x),z_t)
\end{equation}
so that the equilibrium price function $\mathcal{P}^*$ again depends on the entire cross-sectional distribution $G_t$. Solution methods for heterogeneous agent models should be able to handle this generic case. In more general models, other endogenous equilibrium objects besides prices themselves can directly enter decision makers' problems. An example of such ``price-like variables" is the job finding rate in search-and-matching models. In equilibrium, these variables will depend on the distribution $G_t$ just like in \eqref{eq:price_generic}. The vector $p_t$ should therefore be understood as including not only prices themselves but also other price-like variables.

\paragraph{Households and firms with rational expectations forecast prices by forecasting distributions.} To clearly see this implication of rational expectations, consider the special case with two time periods $t=0,1$.

In this case, the households' problem \eqref{eq:household_HA} can be written recursively as follows:
\begin{equation}\label{eq:monster}
\begin{split}
V_0(a,y,G,z) &= \max_{c,n,a'} \ U(c,n) + \beta \mathbb{E}[V_{1}(a',y',G',z')|y,G,z] \quad \mbox{s.t.} \\
c + a' &= w_0^*(G,z) y n + (1+r_0^*(G,z)) a \\[0.5ex]
V_1(a',y',G',z') &= \max_{c',n'} \ U(c',n')  \quad \mbox{s.t.} \quad c' = w_1^*(G',z')y'n' + (1+r_1^*(G',z')) a'.
\end{split}
\end{equation}
Here $V_0$ and $V_1$ are the value functions at times $t=0$ and $t=1$ (which are time-dependent because of the finite horizon). $G'$ is the cross-sectional distribution at time $t=1$ which is induced by households' optimal saving policy $a'=s_0(a,y,G,z)$ and the transition probabilities for labor income $y$, starting from the initial distribution $G$: 
$G' = \mathcal{T}_{s_0} G$ for some operator $\mathcal{T}_{s_0}$ that depends on the function $s_0$.\footnote{In the finite-state variant of Section \ref{sec:not_markov}, the analogue of the operator $\mathcal{T}_{s_0}$ is the big $N \times N$ transition matrix $\mathbf{A}$.}

Importantly, the expectation $\mathbb{E}$ in the first line of equation \eqref{eq:monster} is taken not only over future productivity realizations $z'$ but also over future distributions $G'$. This is because households understand that equilibrium prices at time $t=1$ depend on this distribution -- see $w_1^*(G',z')$ and $r_1^*(G',z')$ in the third line -- and hence so does the future value function $V_1(a',y',G',z')$. Therefore households forecast these prices by forecasting $G'$ in addition to  $z'$.

In terms of computations, the problem is of course the dependence of the value function $V_1(a',y',G',z')$ on the distribution $G'$. Because this distribution is an infinite-dimensional object (or a very high-dimensional one when it is approximated), this leads to an extreme version of the curse of dimensionality. This problem was already noted by \citet{krusell-smith} and \citet{denhaan}. It is also worth noting
%that the dependence of the initial prices $w_0,r_0$ on the distribution is unproblematic: the distribution $G$ is known at $t=0$ and hence so are $w_0,r_0$ in the second line; the problem is instead about future distributions $G'$. Similarly, it is worth noting
that the extreme curse of dimensionality arises exclusively because $G'$ enters decision makers' expectations and \emph{not} because of the distribution evolving over time per se. Computing $G_{1}(a,y)$ given $G_0(a,y)$, a policy function $s_0$, and the productivity shock $z_t$, is computationally straightforward. Heterogeneous-agent models become unrealistically complex only when the distribution enters decision makers' expectations.

The firm's investment decision can be written in a similar fashion and has analogous properties. Just like \eqref{eq:monster} it features future prices $w_1^*(G',z')$ and $r_1^*(G',z')$, a future value function $J_1(k,G',z')$, and an expectation taken over future distributions $G'$.

Equation \eqref{eq:monster} is a two-period version of the ``Master equation" from the Mean Field Games literature \citep{cardaliaguet-delarue-lasry-lions} which has also found its way into economics in recent years \citep[e.g.][]{macro-annual,schaab,bilal,gu-lauriere-merkel-payne}.\footnote{As \citet{sargent-after-lucas} points out, the Master equation is an extension of the ``Big $K$, litte $k$" trick used for solving representative-agent competitive equilibria. Informally, the Master equation is $(k,G(k))$ with individual state $k$ and distribution $G(k)$ and $(k,K)$ is the special case in which all individuals are identical ($G$ is a Dirac point mass at $K$).} To be clear already the work of \citet{krusell-smith} and \citet{denhaan} effectively featured this Master equation, even though they did not use this name.

Using the general notation introduced above with a distribution $G_t(x)$ over a vector of idiosyncratic states $x$ and a vector of aggregate states $z$, this Master equation is a Bellman equation for the value function $V(x,G,z)$ which features the infinite-dimensional state variable $G$. In the MFG literature this Master equation has aptly been nicknamed the ``Monster equation,'' precisely due it being a Bellman equation with an infinite-dimensional state variable. Not surprisingly, solving this Monster equation is computationally extremely challenging -- and often prohibitively so.

In Section \ref{sec:lucas-neutrality}, I cited \citet{lucas-neutrality} as writing: \emph{``Equilibrium prices and quantities will be characterized mathematically as functions defined on the space of possible states of the economy, which are in turn characterized as finite dimensional vectors."} The whole problem is that, in heterogeneous-agent models, these ``states of the economy" are extremely high-dimensional: they are either literally infinite-dimensional cross-sectional distributions or, when the state space is discretized, they are extremely high-dimensional histograms, say $100,000$-dimensional vectors (Section \ref{sec:not_markov} below).

To make matters worse, forecasting future distributions $G'$ requires knowledge not only of the current distribution $G$ but also of all \emph{other} decision makers' optimal choices: the operator $\mathcal{T}_{s_0}$ depends on the saving policy function $s_0(a,y,G,z)$ which summarizes \emph{other} households' saving decisions. Using the notation of Section \ref{sec:not_markov} below, forecasting future distributions requires knowledge of the huge transition matrix $\mathbf{A}$ for the whole economy that summarizes how all \emph{other} decision makers transition across the economy's state space. In the words of \citet{morgenstern}, \emph{``the forward-looking individual must [...] know not only exactly the influence of his own actions on prices, but also that of all other individuals."}

\paragraph{The unrealism of rational expectations.} 
%This paper's main argument is that we should not make our lives so hard. It seems self-evident that real-world households and firms do not forecast prices by forecasting cross-sectional distributions and instead solve simpler problems. Instead of solving ``Monster equations" we should solve the simpler equations corresponding to actual price-forecasting behavior. Below I will spell out some ideas for doing so.
This paper's main argument is that this forecasting behavior implied by rational expectations is extremely implausible. If even our most advanced computational tools struggle with the ``Monster equations," how can we justify the assumption that real-world households and firms solve the associated decision problems? It seems self-evident that households and firms do not forecast prices by forecasting cross-sectional distributions. Instead of solving ``Monster equations" we should replace the rational expectations assumption and solve the simpler equations corresponding to their actual price-forecasting behavior. Below I will spell out some ideas for doing so.

The argument that the rational expectations assumption is often unrealistic is, of course, not new. For example, \citet{manski} writes: \emph{``Suppose that the true state of nature actually is the realization of a random variable distributed $P$. A decision maker attempting to learn $P$ faces the same inferential problems -- identification and induction from finite samples -- that empirical economists confront in their research. Whoever one is, decision maker or empirical economist, the inferences that one can logically draw are determined by the available data and the assumptions that one brings to bear. Empirical economists seldom are able to completely learn objective probability distributions of interest, and they often cannot learn much at all. It therefore seems hopelessly optimistic to suppose that, as a rule, expectations are either literally or approximately rational."}

Similarly \citet{adam-marcet} write: \emph{``The rational expectations hypothesis (REH) places enormous demands on agents' knowledge about how the market works. For most models it implies that agents know exactly what market outcome will be associated with any possible contingency that could arise in the future. This appears utterly unrealistic given that state contingent markets that could provide agents with such detailed information often fail to exist."} Also see the excerpts from \citet{morgenstern} in the introduction and Appendix \ref{sec:morgenstern}.

The argument here is exactly the same but it applies a fortiori because of the model economy's high complexity. Also, recall again that decision makers do not even directly care about the cross-sectional distribution $G$ so it seems particularly questionable that they would solve such complicated dynamic programming problems featuring $G$ as a state variable.

\paragraph{Solution methods for the heterogeneous agent case.} How then do existing numerical methods go about solving this type of problem? There are broadly three sets of solution methods.

The first set of methods assumes that aggregate uncertainty comes in the form of ``MIT shocks" or linearizes with respect to aggregate states. These are the most common methods in the recent heterogeneous-agent literature and, specifically, in most of the HANK literature. For example, \citet{kaplan-moll-violante} use ``MIT shocks" and all of \citet{reiter}, \citet{macro-annual}, \citet{SSJ}, \citet{glawion}, \citet{bilal-goyal}, and \citet{bayer-born-luetticke} use linearization methods.
%\footnote{A few authors also use second-order perturbation methods. See for example \citet{bilal}, \citet{bhandari-bourany-evans-golosov} and \citet{bayer-etal-distributional}.}
In fact, \citet{boppart-krusell-mitman} point out a close connection between the two approaches. With MIT shocks, there is certainty about the path of equilibrium prices. To the same effect, linearization implies certainty \emph{equivalence} about prices. Certainty (equivalence) about equilibrium prices means that these methods completely sidestep the key difficulty of price expectations. The downside of this approach is that such methods are not suitable for answering questions in which aggregate risk and non-linearities play a key role, for example understanding financial crises. The same is true, though to a lesser extent, for other \emph{local} methods like second-order perturbation \citep{bilal,bhandari-bourany-evans-golosov,bayer-etal-distributional,SSJ-2nd-order} which may miss important phenomena like non-linear feedback loops and state spaces with crisis regions -- see Section \ref{sec:global}.%\footnote{\label{foot:global_vs_local}As already noted, what I have in mind are heterogeneous-agent versions of models like \citet{he-krishnamurthy-restud,he-krishnamurthy-AER}, \citet{brunnermeier-sannikov}, \citet{elenev-landvoigt-vanN}, \citet{mendoza}, and \citet{bianchi} which need to be solved using \emph{global} solution methods. A useful prototype for such models is the ``double-well" diffusion process $dX_t = V'(X_t)dt + \sigma dW_t$ where $\sigma$ parameterizes uncertainty, $W_t$ is a standard Brownian motion, and where $V(x)$ is a ``double-well potential" function, e.g. $V(x) = \frac{1}{4}x^4 - \frac{1}{2}x^2$ which has minima at $-1$ and $1$ and a maximum at $0$. See for example \citet[][ch.14]{gardiner} and \citet[][ch.7]{pavliotis}. This process has a bimodal stationary distribution $g(x) \propto \exp(-2 V(x)/\sigma^2)$ and trajectories of $X_t$ spend most time oscillating around the two minima of $V$ while occasionally hopping between these. Perturbation methods around $\sigma=0$ completely miss this type of behavior \citep[e.g.][ch.7.2.4]{gardiner}. Standard references on perturbation methods contain other examples of behavior such methods cannot capture. Also see \citet{farmer-toda} for an example of a representative-agent asset-pricing model in which even a 5th-order perturbation method is highly inaccurate relative to global methods.} 

The second set of methods tackles the global solution of the full rational expectations equilibrium, i.e. of the Master equation. This is where the recent literature has made the most impressive computational advances \citep{schaab,maliar-maliar-winant,azinovic-gaegauf-scheidegger,bilal,gu-lauriere-merkel-payne,gopalakrishna-gu-payne,han-yang-e,huang,proehl}. However, despite these advances, solving the Master equation has remained extremely difficult and this set of methods is what this paper's criticism is aimed at. As stated earlier, in my view, too much intellectual and computational horse power is aimed at solving this unrealistically complex problem.

The third set of methods is that of \citet{krusell-smith} and \citet{denhaan} and variants thereof.\footnote{See for example the continuous-time variants by \citet{FV-hurtado-nuno}  and \citet{lee}.} Compared to solving the full rational expectations equilibrium in which decision makers forecast prices by forecasting distributions, the Krusell-Smith approach instead assumes that they forecast prices by forecasting \emph{moments} of these distributions, say the mean $\bar{a}_t = \int a dG_t(a,y)$ instead of the entire distribution $G_t(a,y)$. Therefore these moments, rather than the entire distribution, become state variables in decision makers' Bellman equations. As \citet{krusell-smith} discuss, this approach has a bounded rationality interpretation: decision makers assume that distributional moments follow a (typically log-linear) perceived law of motion which differs from the objective law of motion. This approach considerably simplifies the computation of heterogeneous agent models with aggregate risk, including in models with aggregate non-linearities. However, just like the full rational expectations approach, the idea that decision makers forecast means, variances, etc of distributions in order to forecast prices lacks realism. In terms of the three criteria spelled out below, I conjecture that this approach would fail on Criterion 2, consistency with empirical evidence.\footnote{I am not aware of any empirical evidence that supports the idea that decision makers forecast, say, the mean of the U.S. wealth distribution to forecast interest rates. \citet{broer-etal-KS} criticize the idea that decision makers use common moments to forecast prices from a theoretical perspective.}

There is one variant of the Krusell-Smith approach which seems to me relatively more promising. This is the variant in which the ``Krusell-Smith moments" are the prices themselves, i.e. decision makers forecast prices directly rather than indirectly via forecasting other moments like means and variances. See for example \citet{lee-wolpin}, \citet{storesletten-telmer-yaron}, \citet{gomes-michaelides}, \citet{favilukis-ludvigson-vanN}, \citet{llull}, \citet{kaplan-mitman-violante}, and \citet{FV-marbet-nuno-rachedi}. I will return to this idea in the next section. Still, I believe that it would be beneficial to develop alternatives to rational expectations about equilibrium prices in a more systematic fashion, in particular taking into account empirical evidence about expectations formation.

\subsection{Taking stock and what next?}
As we saw in Section \ref{sec:roots}, Lucas, Prescott, and their contemporaries aimed to develop operational macroeconomic theories that can be computed, fit to data, and that generate clear policy counterfactuals. In representative-agent models, rational expectations -- though arguably unrealistic -- have achieved exactly this goal.

In heterogeneous agent macroeconomics, however, the rational expectations assumption has the opposite effect, severely hampering the models' operability. It attributes to people an understanding of the price functions that map cross-sectional distributions to equilibrium prices, implying that decisions makers forecast prices by forecasting distributions. This implication is not only unrealistic but also makes such models challenging and costly to solve numerically.

The natural way forward is to replace rational expectations about equilibrium prices in heterogeneous agent models with plausible alternatives. While there are some precedents in the existing literature (most notably the variant of the Krusell-Smith approach discussed above), it is worth to develop such alternatives more systematically and to incorporate key empirical evidence. Doing so promises to ``kill two birds with one stone'': simplifying the computation of heterogeneous-agent models with aggregate risk while, at the same time, making these models more realistic and more likely to generate interesting macroeconomic phenomena.

\section{Three criteria for replacing rational expectations in heterogeneous agent models \label{sec:replace}}

What should replace the assumption of rational expectations about equilibrium prices? I propose three criteria that such alternatives should satisfy. A common element in this class of alternative approaches is that decision makers form expectations about prices directly rather than indirectly by forecasting distributions. While this seems to me the natural solution, any such model of expectations formation necessarily departs from the rational benchmark.\footnote{While all alternatives discussed below drop rational expectations about equilibrium prices, for simplicity, they keep rational expectations about non-equilibrium variables, e.g. idiosyncratic $y_{it}$.} The central question is how to discipline the subjective probability distributions of equilibrium prices, i.e. how to navigate the ``wilderness of non-rational expectations." My three criteria aim to help navigate this wilderness.

\subsection{The natural solution: forecast prices directly \label{sec:direct}}
Recall from Section \ref{sec:HA} that, with rational expectations, households and firms forecast equilibrium prices by forecasting distributions and solve complex Master equations like \eqref{eq:monster}. The most natural alternative is to instead assume that they forecast prices directly in some way.

To flesh this out in more detail, recall the 2-period Master equation \eqref{eq:monster} which features an expectation over the cross-sectional distribution $G'$. The rough idea is that households instead solve 
\begin{equation}\label{eq:natural}
\begin{split}
V_0(a,y,w,r) &= \max_{c,n,a'} \ U(c,n) + \beta \widetilde{\mathbb{E}}[V_{1}(a',y',w',r')| \cdot] \quad \mbox{s.t.} \\
c + a' &= w y n + (1+r) a, \\[0.5ex]
V_1(a',y',w',r') &= \max_{c',n'} \ U(c',n')  \quad \mbox{s.t.} \quad c' = w'y'n' + (1+r') a',
\end{split}
\end{equation}
where the conditional expectation $\widetilde{\mathbb{E}}$ is computed using \emph{some} subjective beliefs about future prices $(w',r')$, i.e. using a conditional probability distribution $\mathbb{P}(w',r'|\cdot )$ that may condition on a number of factors, for example current prices $(w,r)$ or individual factors (so that beliefs are heterogeneous -- see below). Having specified such subjective price beliefs, solving the model is not hard: given beliefs solve for actual prices and quantities such that households and firms maximize and markets clear (a \emph{temporary equilibrium} -- see Section \ref{sec:TE} below).

This is true much more generally. Again using the general notation from above with a distribution $G_t(x)$ over a vector of idiosyncratic states $x$, a vector of aggregate states $z$, and a price vector $p$, all that is needed is a subjective probability distribution over future prices $\mathbb{P}(p'|\cdot)$. With this probability distribution in hand, one can immediately compute expected values $\widetilde{\mathbb{E}}[V(x',z',p')|\cdot]$  and hence solve the Bellman equation for the value function $V(x,z,p)$ as well as solving for the economy's temporary equilibrium given such beliefs. As discussed in Section \ref{sec:HA}, the vector $p$ may also include other ``price-like variables" like the endogenous job finding rate in search-and-matching models. The natural assumption is that decision makers also forecast these other variables directly.

As an aside, it is useful to ask: why can we not just write a Bellman equation \eqref{eq:natural} with prices $p$ as state variables in the rational expectations case? That is, why can we not just write \eqref{eq:natural} but with the subjective expectation $\widetilde{\mathbb{E}}$ replaced by the rational expectation $\mathbb{E}$? The short answer is: because prices $p$ do not follow a Markov process, a point I flesh out in detail in Section \ref{sec:not_markov} below.

The focus on a low-dimensional vector of equilibrium variables links the proposed approach to the ``sequence space" methods of \citet{SSJ}. The difference is that my aim is to handle stochastic price sequences outside steady-state neighborhoods, i.e. a global rather than local solution. Handling such sequences in a recursive fashion requires departing from rational expectations precisely because prices do not have the Markov property.

It is also worth noting again that the proposed approach in which decision makers forecast prices directly differs not only from the rational-expectations approach (in which decision makers forecast prices indirectly by forecasting cross-sectional distribution) but also from the Krusell-Smith approach (in which decision makers forecast prices indirectly via forecasting moments of the distribution). The one exception is the more promising variant of the Krusell-Smith approach in which these moments are the prices themselves (Section \ref{sec:HA_forecasting}).

Finally, I wrote the value functions in \eqref{eq:natural} as featuring only the prices $p=(w,r)$ in addition to the idiosyncratic states $x=(a,y)$. This is more restrictive than necessary. Specifically, the subjective beliefs $\mathbb{P}(p'|\cdot )$ could also condition on (a small number of) other state variables that forecast prices. For example, the percentage of households with mortgages may forecast house prices. The vector of additional states in the value function would then also include these other variables. Decision makers forecasting prices directly means that this vector of forecasting states contains all payoff-relevant prices (but it may contain other variables as well).

\subsection{The challenge: disciplining price expectations}
The challenge is, of course, how to discipline the subjective probability distributions for future prices $\mathbb{P}(p'|\cdot)$ that are then used to compute price expectations $\widetilde{\mathbb{E}}[V(x',z',p')|\cdot]$. This is the challenge of navigating the ``wilderness of non-rational expectations." As \citet{sargent-presidential} put it: \emph{``There is such a bewildering variety of ways to imagine discrepancies between objective and subjective distributions,"} followed by a pointed footnote \emph{``There is an infinite number of ways to be wrong, but only one way to be correct."}

To help navigate this wilderness, I next outline three criteria that alternative approaches should fulfil in my view. Sargent's piece advocated \emph{``cautious modifications of rational expectations theories [...] to retain the discipline of rational expectations econometrics."} The three criteria below share this goal while, at the same time, aiming to improve the realism and computation of heterogeneous agent models.

\subsection{Three criteria for price expectations \label{sec:criteria}}
The following three criteria are natural requirements on subjective price beliefs $\mathbb{P}$. I first list the three criteria and then discuss each of them in turn.
\begin{enumerate}
\item Computational tractability
\item Consistency with empirical evidence 
\item Endogeneity of beliefs to model reality (Lucas critique)
%Some consistency between beliefs and model reality with beliefs responding to policy (Lucas critique)
\end{enumerate}
The challenge posed in this paper is to develop alternatives to rational expectations about equilibrium prices that fulfil these three criteria. The hope is that these criteria \emph{in combination} will provide useful guideposts for navigating the wilderness of non-rational expectations and to considerably narrow down the list of candidate subjective price beliefs $\mathbb{P}$.

\paragraph{Criterion 1: Computational tractability.} The first criterion is that the candidate alternative approach should actually simplify the computational solution of heterogeneous agent models with aggregate risk (relative to the Master equation approach), i.e. it should make these models ``operational." More precisely, the alternative approach should simplify the problem of individual decision makers and result in Bellman equations with much lower-dimensional state variables.

This criterion actually has quite a bit of bite and rules out (standard versions of) several popular expectations models. In particular, it eliminates any approach that begins with rational expectations as a special case and then derives the subjective probability distribution $\mathbb{P}$ by merely ``twisting" this objective distribution. This is because the rational expectations case already suffers from the extreme curse of dimensionality discussed above, and any generalization is inevitably at least as complex.\footnote{Another class of models that does not satisfy my Criterion 1 for essentially the same reasons is the Limited Information Rational Expectations (LIRE) class, i.e. models that relax only the FI part but not the RE part of Full Information Rational Expectations (FIRE). LIRE includes noisy rational expectations \citep{lucas-neutrality} and rational inattention \citep{sims-RI,mackowiak-matejka-wiederholt}.}

More precisely, several models of non-rational expectations in the literature index subjective beliefs by a parameter $\theta$, i.e. $\mathbb{P}^\theta$, with $\theta = 0$ corresponding to rational expectations, so that $\mathbb{P}^{\theta=0}$ equals the objective distribution $\mathbb{P}^{RE}$. These include the standard formulations of diagnostic expectations \citep{bordalo-gennaioli-shleifer,bordalo-gennaioli-shleifer-JEP} and cognitive discounting  \citep{gabaix-sparsity,gabaix-NK}. For example, the standard version of diagnostic expectations would postulate price beliefs $\mathbb{P}^\theta$ as
$$\mathbb{P}^\theta(p|D) \propto \mathbb{P}^{RE}(p|D) \left[\frac{\mathbb{P}^{RE}(p|D)}{\mathbb{P}^{RE}(p|-D)}\right]^\theta, $$
where $D$ is recently observed data, $-D$ is data in a relevant comparison group, and where $\theta \geq 0$. When $\theta=0$ the special case of rational expectations obtains. Similarly, in the cognitive discounting approach of \citet{gabaix-NK}, subjective expectations $\widetilde{\mathbb{E}}$ are defined from the corresponding rational expectations $\mathbb{E}^{RE}$ as $\widetilde{\mathbb{E}}[p] = \widetilde{\mathbb{E}}[\bar{p} + \widehat{p}] = \bar{p} + \bar{m} \mathbb{E}^{RE}[\widehat{p}]$, where $\bar{p}$ is a default value (e.g. the steady state), $\widehat{p}$ the deviation from this default value, $\bar{m} \in [0,1]$ a ``cognitive discounting" parameter, and where $\bar{m}=1$ corresponds to rational expectations; equivalently, $\theta=1-\bar{m}$ with rational case $\theta=0$. A third example is the recursive competitive equilibrium with behavioral agents in Section 5.3 of \citet{gabaix-JEEA}. All of these models do not satisfy my first criterion.\footnote{That being said, variants of the standard formulations may well satisfy Criterion 1. For example, both diagnostic expectations and cognitive discounting are specified \emph{relative} to some other beliefs and these other beliefs need not be rational \citep{hajdini}. In the equation above $\mathbb{P}^{RE}$ could be replaced by some alternative baseline $\mathbb{P}$. But this leaves open the question what that alternative baseline $\mathbb{P}$ should be. Similarly, one can imagine a variant of the \citet{gabaix-JEEA} recursive competitive equilibrium with behavioral agents in which decision makers use a simplified model for equilibrium prices (e.g. an autoregressive process) which endogenously becomes more sophisticated when the stakes increase or the attention cost falls.}

Put differently, for the purposes of this paper, models of non-rational expectations that require decision makers to be able to compute the special case of rational expectations defeat the point of departing from rational expectations in the first place. This is true not just from this paper's pragmatic, computational viewpoint but arguably also from that of the behavioral macroeconomics literature: while such models have been extremely useful for exploring the implications of deviating from the rational expectations hypothesis (see e.g. the papers above), they are an intermediate step toward more satisfactory theories of bounded rationality that do not require decision makers to access the full rational expectations solution.

%requires people to hold a whole distribution of inequality in their heads

\paragraph{Criterion 2: Consistency with empirical evidence.} The second criterion is that the candidate alternative approach should be consistent with established findings from the empirical expectations literature. Over the last twenty years, this literature has assembled a large body of evidendence, in particular by measuring expectations through surveys \citep[e.g.][]{manski,mankiw-reis-wolfers,armantier-etal,manski-macroannual,weber-dacunto-gorodnichenko-coibion,dacunto-weber}. There are now new surveys like the New York Fed's Survey of Consumer Expectations (SCE) and even a Handbook of Economic Expectations \citep{handbook-expectations}.

This literature has documented a number of empirical patterns, some of which I summarize in Section \ref{sec:survey} below. While survey expectations data need to be interpreted with care (e.g. due to concerns about systematic noise), candidate alternative approaches should take on board this empirical evidence. For example, one important finding in the survey expectations literature is the vast observed heterogeneity of measured beliefs which stands in contrast to the ``communism'' of rational expectations \citep{sargent-palgrave}, i.e. that all decision makers share the same beliefs. This suggests developing alternatives to rational expectations in which subjective price beliefs differ in the population, that is, modeling subjective probability distributions $\mathbb{P}_i$ where $i$ indexes individual households or firms. It is worth noting that the tools of heterogeneous agent macroeconomics are well-suited to modeling belief heterogeneity: after all, modeling and ``carrying around'' cross-sectional distributions is our bread and butter -- so why not also add a distribution of beliefs to these models?\footnote{\citet{guerreiro} takes a step in this direction but linearizes his model. This simplifies the analysis because he only has to track a distribution of mean beliefs (expectations) across (a relatively small number of) household groups. As noted elsewhere, linearized models are not suitable for modeling phenomena like financial crises. For other dynamic models of belief heterogeneity -- though many of them in restrictive linear-Gaussian setups -- see for example \citet{mankiw-reis},  \citet{woodford_2001,woodford-finite}, \citet{giannitsarou}, \citet{carroll}, \citet{lorenzoni}, \citet{angeletos_lao_2013}, \citet{branch-mcgough-handbook}, \citet{chahrour-gaballo}, \citet{carroll-etal-sticky}, \citet{bouchaud-farmer}, and \citet{straub-ulbricht}.}

In economics there is often a trade-off between realism and (computational) simplicity: the real world is a complex place so more of one usually means less of the other. One may have therefore conjectured that Criteria 1 and 2 are in conflict with each other. But this is likely not the case. Instead realism and simplicity may go hand in hand in this case, offering an opportunity to kill two birds with one stone.

There are three other types of empirical evidence that models employing the candidate alternative approach may want to match. First, evidence on the pass-through from subjective beliefs to decision makers' choices. Second, evidence on the pass-through from shocks and policies to subjective beliefs (Section \ref{sec:survey}). Third, the time-series properties of observed prices and price-like variables (see the discussion of Criterion 3).

\paragraph{Criterion 3: Endogeneity of beliefs to model reality (Lucas critique).} The third criterion is that the candidate alternative approach should imply price beliefs that are endogenous to the behavior of actual equilibrium prices. It is useful to split Criterion 3 into two sub-criteria:

\medskip

\noindent \textbf{Criterion 3a:} In stationary environments with a sufficiently long time series for prices, subjective price beliefs should be approximately consistent with actual (model) equilibrium prices.

\medskip

\noindent \textbf{Criterion 3b:} In all environments, including unfamiliar and non-stationary ones, price beliefs should respond to model reality, specifically to policy changes (Lucas critique).

\medskip

I now discuss these two sub-criteria in turn. Criterion 3a is concerned with familiar environments in which equilibrium prices follow ``regular patterns." In such situations, price beliefs should not be ``too far" from the behavior of actual equilibrium prices. Denoting the subjective probability distribution of prices by $\mathbb{P}(p|\cdot)$ and the objective price distribution by $\mathbb{P}^{obj}(p|\cdot)$, we should have
\begin{equation}\label{eq:distance}
||\mathbb{P}(p|\cdot) - \mathbb{P}^{obj}(p|\cdot)|| \leq \varepsilon
\end{equation}
for some distance metric (norm) $||\cdot||$ and some ``small" scalar $\varepsilon \geq 0$. While rational expectations imposes exact consistency between subjective and objective distributions, $\mathbb{P}^{RE}(p|\cdot) = \mathbb{P}^{obj}(p|\cdot)$ identically, Criterion 3a imposes approximate consistency. 

The rationale for Criterion 3a is already in the excerpts from \citet{lucas-methods} and \citet{lucas-nobel} in Section \ref{sec:bob}, in particular that \emph{``the implied behavior of actual equilibrium prices [should not be] grossly inconsistent with the price expectations that the theory imputes to individual agents."} Without this criterion, alternative approaches would go full circle back to the 1960s. The argument here is that weaker restrictions than rational expectations can also deliver on Lucas' desiderata. This criterion is nothing new and has been advocated by other proponents of (cautious) departures from rational expectations. For example, \citet{woodford-AR} writes: \emph{``It makes sense to assume that expectations should not be completely arbitrary and have no relation to the kind of world in which the agents live [...] We should therefore like to replace the RE hypothesis by some weaker restriction that nonetheless implies a substantial degree of conformity between people's beliefs and reality."} Also see  \citet{sargent-bounded}, \citet{sargent-conquest}, and \citet{ilut-schneider-AER} who impose an approximate consistency criterion similar in spirit to \eqref{eq:distance} for worst- and best-case means in a model with ambiguity aversion.\footnote{\citet{sargent-encyclopedia} has invoked Abraham Lincoln's famous phrase to describe the philosophy behind rational expectations: \emph{``You can fool some of the people all of the time, and all of the people some of the time, but you cannot fool all of the people all of the time."} Criterion 3 is designed to ensure that this remains true. Note that Lincoln's quote explicitly allows for people being fooled \emph{some} of the time and therefore arguably aligns more closely with the types of beliefs advocated here than with the strict rational expectations assumption itself.}

% }
%\footnote{In Ilut and Schneider's model, each decision maker entertains a time-varying set of beliefs about an ambiguous variable $z_{t}$ parameterized by a set of mean beliefs in an interval $[-a_t,a_t]$ where $-a_t$ and $a_t$ are the time-varying worst- and best-case means so that $a_t$ parameterizes ambiguity. They impose a model consistency criterion between beliefs and the true data generating process in the form of a bound on ambiguity such that  XXX $a_t \leq 2\sqrt{Var(z_t)}$, meaning that decision makers do not entertain forecasts $a_t$ that are too extreme }

The role of Criterion 3b is that Criterion 3a may be too strong in many contexts, specifically in ``unfamiliar" situations that have ``not been seen before." Extreme cases include the Covid pandemic, the 9/11 terrorist attacks, or the 2000s housing boom, but many less dramatic real-world events also challenge the assumption of stationarity. In such contexts, even the best econometricians would struggle to closely approximate the true data-generating process for equilibrium prices. Therefore, price beliefs of decision makers in our models may differ from the true process by a wider margin. This idea is consistent with experimental evidence on reasoning in games that beliefs differ more from rational beliefs in new environments than in familiar ones \citep[e.g.][]{nagel-guessing-games}. Similarly, \citet{angeletos-lian-dampening} write: \emph{``The assumption that the subjective statistical model in people's mind is the same as the objective truth, is hard to defend in the context of unusual circumstances, such as the Great Recession and the ZLB on monetary policy."} \citet{bouchaud-farmer} argue that the rational expectations assumption is difficult to justify even beyond such unusual circumstances because most real-world stochastic processes are neither stationary nor ergodic.

Nevertheless, while such reasoning supports relaxing Criterion 3a, price beliefs should not be completely exogenous to the model. Because equilibrium prices are themselves endogenous, beliefs about them must be modeled as at least partially endogenous as well. This is the content of Criterion 3b. For example, when conducting policy counterfactuals, one cannot simply plug empirical survey expectations into the model and compute the corresponding temporary equilibria. Instead, price beliefs need to be modeled as endogenous. The behavior of actual equilibrium prices depends on price beliefs and, conversely, price beliefs depend on the behavior of actual prices -- a fixed point problem. Concretely, we need to model how expectations respond to policy changes. If beliefs were completely unresponsive to policy, the model would be subject to the \citet{lucas-critique} critique.

For both Criteria 3a and 3b, an important question is how to evaluate the size of ``mistakes", i.e. what distance metric between subjective and objective price distributions to use. A natural approach is to use a welfare-based metric, i.e. to impose that mistakes should not be too costly in terms of welfare. Also this idea is not new. \citet{lucas-prescott} already criticize models in which the difference between price forecasts and actual prices is \emph{``persistent [and] costly to forecasters"} in the passage quoted in Section \ref{sec:lucas-prescott}. The idea to judge mistakes in terms of welfare costs similarly lies at the heart of models of both rational and behavioral inattention (but see the discussion surrounding Criterion 2).

Finally, if we want to impose some consistency between beliefs and model reality (Criterion 3) while, at the same time, imposing consistency with empirical evidence on beliefs about real-world prices (Criterion 2), there should also be some consistency between model-generated and real-world prices. That is, our heterogeneous-agent model should generate empirically reasonable time-series behavior for equilibrium prices.

To deliver on both parts of Criterion 3, an obvious approach is to model some form of learning. Models of learning have been a key focus of the existing literature and will occupy a large part of the discussion of promising directions in Section \ref{sec:directions} below.

\subsection{The need for global and recursive solution methods \label{sec:global}}
In addition to these three criteria for the subjective price beliefs $\mathbb{P}$, an important practical requirement concerns the class of models that the candidate approach can be applied to: non-linear models that are solved using global, recursive solution methods.

\paragraph{Why global solution methods?}
As stated above, one of the ultimate goals of developing alternatives models of price beliefs is to be able to consider questions in which aggregate risk and aggregate non-linearities play a key role. Therefore, the candidate alternative approach should be applicable in fully non-linear models. In contrast, much of the theoretical expectations literature has considered linear (or linearized) models, likely because it means that standard techniques like the Kalman filter can be applied. This would need to change.

As already noted in the introduction, what I have in mind are heterogeneous-agent versions of models like \citet{he-krishnamurthy-restud,he-krishnamurthy-AER}, \citet{brunnermeier-sannikov}, \citet{mendoza}, \citet{bianchi}, \citet{elenev-landvoigt-vanN}, and \citet{krishnamurthy-li} which feature phenomena like non-linear feedback loops, state spaces with crisis regions, or bimodal ergodic distributions for the aggregate economy. Such models need to be solved using global solution methods because local perturbation methods fail to capture their behavior.

A useful prototype for such models is the scalar ``double-well" or ``bistable" diffusion process in Appendix \ref{app:global} which features a parameter $\sigma$ that parameterizes uncertainty. Without uncertainty $\sigma=0$ there are three steady states, two of them stable and one unstable. With uncertainty $\sigma>0$, the process has a bimodal stationary distribution. Trajectories spend most time oscillating around the two stable steady states while occasionally hopping between them. Starting from the high steady state, ``the economy" stays close to it most of the time but may be thrown into a ``crisis" (low steady state) and get stuck there for a while before ultimately recovering. Perturbation methods around $\sigma=0$ completely miss this type of behavior \citep[e.g.][ch.7.2.4]{gardiner}. Standard references on perturbation methods also contain other examples of behavior such methods cannot capture.

Finally, even with less complicated dynamics, perturbation methods may be quantitatively off. See \citet{farmer-toda} for an example of a representative-agent asset-pricing model in which even a 5th-order perturbation method is highly inaccurate relative to global methods.

\paragraph{Why recursive methods?} Similarly, the alternative approach must be compatible with recursive solution methods for household and firm optimization problems (Bellman equations). These lie at the heart of solution methods for current heterogeneous-agent models for two reasons. First, recursive methods naturally handle micro non-linearities such as borrowing constraints. Second, rather than solving a separate optimization problem for each of a large number of heterogeneous decision makers (i.e. millions of optimization problems), recursive methods often require solving only a single Bellman equation, with the resulting policy function summarizing the actions of (ex post) heterogeneous agents across the state space.

%
%Similarly to the three criteria for subjective beliefs above, these two additional practical requirements should hopefully serve to narrow down the list of applicable approaches.

\subsection{With rational expectations, why can we not carry prices as state variables? \label{sec:not_markov}}

Before discussing promising directions, I briefly revisit the question: with rational expectations, why can we not just write a Bellman equation with prices $p$ as state variables? Why can we not write \eqref{eq:natural} but with the subjective expectation $\widetilde{\mathbb{E}}$ replaced by the rational expectation $\mathbb{E}$?

The short answer is that equilibrium prices do not satisfy the Markov property. As a reminder, a stochastic process has the Markov property if the probability distribution of future states depends only on the present state. For a discrete-time process $\{p_t\}$, the Markov property would mean that one can write the conditional distribution of $p_{t+1}$ as a function of $p_t$ only: $\Pr(p_{t+1}|\cdot) = \Pr(p_{t+1}|p_t)$.

To see that prices do not satisfy this property, it is useful to consider a variant of the general model in Section \ref{sec:HA_forecasting} with a distribution over the vector of idiosyncratic states $x$, aggregate states $z$, and a price vector $p$, but with a finite state space for the idiosyncratic states, i.e. $x$ can take only $N<\infty$ possible values, $x \in \{x_1,...,x_N\}$, say $N=100,000$. In this case, the distribution $G_t$ is simply an $N$-dimensional vector $\mathbf{G}_t=[G_{1,t},...,G_{N,t}]^{\rm T}$ and it is convenient to work with the corresponding density $\mathbf{g}_t=[g_{1,t},...,g_{N,t}]^{\rm T}$ which is simply the ``histogram" collecting the fraction of agents at each point of the state space.

As above, the vector of equilibrium prices is a function of the entire distribution:
\begin{equation}\label{eq:price_function_finite}
p_t = \mathcal{P}^*(\mathbf{g}_t,z_t),
\end{equation}
I use regular notation for the vector $p_t$ but bold-face notation for the vector $\mathbf{g}_t$ to signify that $p_t$ is typically much lower-dimensional than $\mathbf{g}_t$, say $p_t \in \mathbb{R}^2$ versus $\mathbf{g}_t \in \mathbb{R}^{100,000}$.

To see why prices do not satisfy the Markov property, consider the evolution of $\mathbf{g}_t$. This density evolves according to a Chapman-Kolmogorov equation (the discrete-time analogue of a Kolmogorov-Forward equation):
\begin{equation}\label{eq:KF}
\mathbf{g}_{t+1} = \mathbf{A}(\mathbf{g}_t,z_t,p_t)^{\rm T} \mathbf{g}_{t} 
\end{equation}
where $\mathbf{A}(\mathbf{g}_t,z_t,p_t)$ is the transition matrix of the vector of idiosyncratic states $x$ (with $x$ evolving according to the optimal policies from agents' Bellman equation).\footnote{See \citet{AHLLM} for a continuous-time analogue and an explanation why the Kolmogorov Forward (KF) equation features the transpose of the transition matrix. Intuitively, the transition matrix $\mathbf{A}$ answers the question ``where in the state space are people going?" but the KF equation asks ``where in the state space are people coming from?" -- hence the transpose.} Note that this is a large matrix of dimension $N \times N$, say with $N=100,000$, that summarizes how all individual decision makers transition across the economy's state space. Importantly, after substituting in for $p_t$ from \eqref{eq:price_function_finite}, this equation defines a \emph{Markov process for the distribution $\mathbf{g}_t$ and aggregate shock $z_t$}: we have $\Pr(\mathbf{g}_{t+1},z_{t+1}|\cdot) = \Pr(\mathbf{g}_{t+1},z_{t+1}|\mathbf{g}_t,z_t)$.

Summarizing, prices $p_t$ depend on the distribution and aggregate shock $(\mathbf{g}_t,z_t)$ according to the equilibrium price function $\mathcal{P}^*$ in \eqref{eq:price_function_finite} and $(\mathbf{g}_t,z_t)$ follows a Markov process. This immediately implies that prices $p_t$ themselves do \emph{not} follow a Markov process. Instead, the stochastic process for $p_t$ is extremely complicated. More precisely, the conditional distribution of future prices satisfies $\Pr(p_{t+1}|\cdot) = \Pr(p_{t+1}|\mathbf{g}_t,z_t)$ rather than $\Pr(p_{t+1}|\cdot) = \Pr(p_{t+1}|p_t)$ meaning that the probability distribution of future prices depends on the entire cross-sectional distribution $\mathbf{g}_t$. Therefore, one cannot simply write a Bellman equation \eqref{eq:natural} with prices $p$ as state variables. Instead, decision makers with rational expectations forecast the Markov state $(\mathbf{g}_t,z_t)$ in order to forecast the non-Markovian prices $p_t$.

Another perspective comes from the analogy with a \emph{Hidden Markov Model} (HMM): the distribution $\mathbf{g}_t$ is a high-dimensional latent state satisfying the Markov property, while $p_t$ is a low-dimensional observation derived from it. Since $p_t$ alone is not Markov, forecasting it requires forecasting the high-dimensional Markov state. Recognizing this structure in heterogeneous-agent models could guide alternatives to rational expectations. For instance, if decision-makers track only prices rather than the full distribution (as in Section \ref{sec:direct}), their optimization problems become \emph{Partially Observable Markov Decision Processes} (POMDPs), i.e., controlled HMMs.

\section{Some promising directions \label{sec:directions}}
In the final section I discuss some potentially promising directions for developing alternative models of price beliefs satisfying my three criteria. These include temporary equilibrium and internal rationality (section \ref{sec:TE}), incorporating survey expectations (section \ref{sec:survey}), least-squares learning and stochastic approximation (section \ref{sec:LSL}), reinforcement learning (section \ref{sec:RL}), and heuristics and simple models (section \ref{sec:simple}). The discussion here is purposely selective. For more complete treatments, see the surveys and books referenced in the introduction.

\subsection{Temporary Equilibrium and Internal Rationality \label{sec:TE}}
The idea of ``temporary equilibrium" is fundamental for developing alternatives to rational expectations about equilibrium prices: all approaches in which decision makers forecast prices directly rather than indirectly via forecasting distributions (Section \ref{sec:direct}) will necessarily involve computing such temporary equilibria as an intermediate step.

\textbf{Definition:} \emph{Temporary equilibrium} at a particular time $t$ is defined as allocations and prices such that (i) households and firms optimize \emph{given expectations of future variables} (including future prices) that are specified in the model but that are \emph{not necessarily rational}, (ii) markets clear at time $t$.

This idea was originally developed contemporaneously by \citet{hicks-value} and \citet{lindahl}, has been further developed by \citet{grandmont,grandmont-palgrave}, and has found a number of applications in the more recent literature.\footnote{Including \citet{woodford-AR}, \citet{piazzesi-schneider-handbook}, \citet{bossaerts-etal-lucas-lab}, \citet{garciaschmidt-woodford}, \citet{farhi-werning-level-k}, and \citet{werning-expectations}. Although they do not connect to the temporary equilibrium idea, another application is the demand-system asset-pricing approach of \citet{koijen-yogo} which replaces investor's beliefs about high-dimensional equilibrium asset returns with a low-dimensional estimated factor model.} Rational expectations equilibrium is the special case of temporary equilibrium in which expectations are rational (model-consistent).\footnote{\citet[][Chapters IX.7 and X.2]{hicks-value} differentiates ``temporary equilibrium" from another notion of equilibrium he calls ``equilibrium over time" which bears some resemblance to a rational-expectations equilibrium. He explains the difference in a dynamic economy with time intervals lasting one week: \emph{``The wider sense of Equilibrium -- Equilibrium over Time, as we may call it, to distinguish it from the Temporary Equilibrium which must rule within any current week -- suggests itself when we start to compare the price-situations at any two dates. [...]"} Also see \citet{hicks-formation} who defines temporary equilibrium as \emph{``a momentary market equilibrium in which price-expectations are taken as data."}}

The approach outlined in Section \ref{sec:direct} was exactly one of temporary equilibrium: given specified beliefs $\mathbb{P}(p'|\cdot)$, compute value functions $V(x,z,p)$ and associated optimal policies, then solve for equilibrium prices that clear markets. It is important to note that computation of temporary equilibria given price beliefs is relatively straightforward with modest computational costs. In particular, in a temporary equilibrium, there is no fixed point between price beliefs and actual equilibrium prices. Therefore, its computation is only slightly more difficult than the computation of a stationary equilibrium in a heterogeneous-agent model \citep[as in][]{aiyagari}. In this regard, it is also worth restating that tracking the distribution $G_t(x)$ over time, i.e. computing $G_{t+1}(x)$ given $G_t(x)$, a policy function for the evolution of $x$, and the aggregate shock $z_t$ is computationally straightforward. For example, in the finite-state-space variant of Section \ref{sec:not_markov} tracking the vector $\mathbf{g}_t$ simply means running the difference equation \eqref{eq:KF} forward in time. Heterogeneous-agent models become unrealistically complex only when the distribution enters decision makers' expectations.

While computing temporary equilibria is relatively straightforward, to fulfil Criterion 3 above, this can only be an intermediate step toward solving the bigger fixed point problem in which the behavior of actual equilibrium prices feeds back into endogenous price beliefs.

\paragraph{Internal Rationality.} When defining temporary equilibria, it is important to derive individual policy functions from intertemporal decision problems with dynamically consistent subjective beliefs about equilibrium prices. This links temporary equilibrium approaches to the concept of \emph{internal rationality} proposed by \citet{adam-marcet}. 

Adam and Marcet distinguish between variables under decision makers' control and those that are beyond their control or ``external," with equilibrium prices being the prime example of such external variables. They argue that the correct way of relaxing rational expectations is to start with a well-defined system of (non-rational) subjective beliefs about these external variables and to then derive decision makers' policy functions from their intertemporal problems given such beliefs. They term this approach \emph{internal rationality} to emphasize that decision makers optimize but without being ``externally rational," i.e. without knowing the objective probability distributions of external variables. Internal rationality contrasts with an alternative approach that starts from individual optimality conditions derived under rational expectations and then simply replaces the rational expectations operators with some other expectations, an approach that can result in inconsistencies \citep{preston}.

The temporary equilibria discussed above should thus be ``Internally Rational Expectations Equilibria" in Adam and Marcet's language. Starting from Bellman equations with specified subjective beliefs will automatically result in internal rationality. Because this is the natural solution method for heterogeneous agent models, the inconsistencies just discussed should not be a concern in practice.

\paragraph{Self-Confirming Equilibrium and Friends.} The economics literature has proposed various concepts of ``non-rational expectations equilibrium" or ``misspecification equilibrium" in which actual equilibrium outcomes are statistically consistent with decision makers' beliefs (i.e. these beliefs are not disappointed) but with  weaker or different consistency requirements than under rational expectations. These alternative equilibrium concepts may be useful building blocks for fulfiling Criterion 3a (``approximate consistency between beliefs and model reality"). In contrast to temporary equilibria in which beliefs are exogenously specified, in these alternative equilibria,  beliefs are endogenous and are the fixed point of some mapping from a (potentially misspecified) perceived law of motion to an approximate actual law of motion.

Appendix \ref{app:equilibria} summarizes a number of these, specifically: self-confirming equilibrium (SCE), restricted perceptions equilibrium (RPE), and consistent expectations equilibrium (CEE). The appendix also puts these equilibrium concepts in relation to temporary equilibrium (TE), internally rational expectations equilibrium (IREE), and  rational expectations equilibrium (REE). As explained there, the relation between the various equilibrium concepts  can be summarized as follows:
\begin{equation}\label{eq:equilibria}
\{REE,RPE,CEE\} \subset SCE \subset IREE \subset TE,
\end{equation}
where $\subset$ means ``is a special case of."

\subsection{Survey Expectations and Hypothetical Vignettes \label{sec:survey}}
As already discussed, to satisfy Criterion 2 (``consistency with empirical evidence"), alternative approaches should incorporate the findings from the large empirical literature on expectations formation including evidence from survey expectations.

This literature has documented a number of empirical patterns that challenge the rational expectations hypothesis. As already noted, the most obvious of these is the pervasive heterogeneity of subjective beliefs or ``belief disagreement." While some of this disagreement may reflect noise (see below), this finding stands in contrast to the ``communism of beliefs" of rational expectations. See for example \citet{mankiw-reis-wolfers}, \citet{malmendier-nagel-depression-babies}, \citet{malmendier-nagel-inflation}, \citet{giglio-etal-fivefacts}, and  \citet{fofana-patzelt-reis}. Similarly, a large literature has documented overreaction of individual-level forecasts to idiosyncratic news and underreaction of average forecasts to aggregate news \citep{coibion-gorodnichenko-JPE,coibion-gorodnichenko-AER,bordalo-gennaioli-shleifer-JEP}. An example of another interesting finding is ``cross-domain extrapolation''.\footnote{For example, \citet{cenzon} documents that people experiencing credit rejections become pessimistic not only about credit markets but also about inflation, unemployment, and stock prices and \citet{taubinsky-butera-saccarola-lian} argue that inflation expectations react excessively to household-level income shocks. Also see \citet{bordalo-etal-imagining}. This evidence, as well as the evidence on subjective earnings risk in the next paragraph, is also of independent interest for heterogeneous-agent modeling, for example it could have important implications for the distribution of wealth.}

To empirically discipline subjective beliefs about equilibrium prices, survey evidence on the relevant expectations is needed. These will differ according to the particular model being considered. In the simple heterogeneous-agent model in Section \ref{sec:HA} one would need survey expectations of future (real) wages and interest rates. In this regard measures of subjective earnings risk are an important piece of evidence, see for example \citet{dominitz-manski}, \citet{mueller-spinnewijn}, \citet{caplin-etal-subjective}, and \citet{balleer-duernecker-forstner-goensch}.

Additionally, one would really need measures of these expectations \emph{conditional} on the state of the aggregate economy (recessions vs booms) and on different policies (policy counterfactuals). Put differently, one would like evidence on the ``subjective models" (perceived laws of motion) that decision makers use to form expectations. One way of eliciting such conditional beliefs using surveys is to use hypothetical vignettes \citep{haaland-roth-wohlfart,andre-etal-subjective-models}.

An important concern about survey expectations is that a substantial fraction of reported beliefs may reflect ``noise", some of it classical measurement error but some of it also systematic error due to respondents' cluelessness about the object they are asked to forecast. A common finding in the literature is that reported subjective probabilities are compressed toward 50:50 relative to true objective probabilities, i.e. survey respondents overstate the probability of unlikely events and understate the probability of likely events \citep[e.g.][]{viscusi,fischhoff-bruinedebruin,enke-graeber}. \citet{enke-graeber} elicit survey respondents' subjective beliefs and additionally ask respondents how certain they are about these stated beliefs (``cognitive uncertainty"). They find that more cognitively uncertain respondents have more compressed beliefs, i.e. that cognitive uncertainty systematically distorts stated beliefs. Therefore heterogeneity in survey expectations may partly reflect heterogeneity in ``cognitive noise" rather than true beliefs.\footnote{Put differently, the concern about systematic distortions is that cognitively uncertain respondents may appear either very optimistic or very pessimistic depending on the particular survey question being asked. For example, asking ``What is the probability that inflation will exceed 10\% over the next year?" (a highly unlikely event), will lead cognitively uncertain households to state a high probability unreasonably close to 50:50 and these households would thus appear very pessimistic about inflation. Conversely, asking ``What is the percent chance that inflation will exceed 0\%?" (a very likely event), will make them appear very optimistic.} This concern about systematic noise in survey expectations is not an argument against the use of such data; rather it calls for interpreting such data with care, particularly in contexts in which cognitive noise may be important, for example when households are asked about objects that are not relevant to them on a daily basis. Another important piece of the puzzle is the pass-through from survey beliefs to actions, see \citet{giglio-etal-fivefacts}, \citet{charles-frydman-kilic}, and \citet{yang}.

While the present section focussed on household expectations, firm expectations about the conditions they operate in are, of course, equally important. See for example 
\citet{bachmann-elstner-sims}, \citet{candia-coibion-gorodnichenko} and \citet{bloom-codreanu-fletcher}.

\paragraph{Temporary Equilibrium with Measured Expectations.} A particularly interesting approach is to compute temporary equilibria in which beliefs are disciplined with survey expectations. This  ``temporary equilibrium with measured expectations'' approach is described in \citet{piazzesi-schneider-handbook} and in Section 2.4 of \citet[][written by Piazzesi]{brunnermeier-piazzesi-etal-RFS}. See \citet{landvoigt-piazzesi-schneider}, \citet{leombroni-piazzesi-rogers-schneider}, \citet{arondine-beutel-piazzesi-schneider}, \citet{bardoczy-guerreiro}, and \citet{ludwig-mankart-quintana-wiederholt} for recent applications.

\citet{roth-wiederholt-wohlfart} extend this approach to computing policy counterfactuals by 
means of measured expectations under different counterfactuals, elicited using hypothetical vignettes. While this approach could, in principle, be used to address the critique that beliefs should respond to policy (Lucas critique), a downside is that a new vignette question would need to be fielded for every new policy counterfactual. The more standard alternative approach is to model expectations in a way that incorporates the relevant empirical evidence (e.g. from survey expectations). Vignette proponents argue in favor of their approach by pointing to the falling cost of conducting surveys and economists' disagreement about modeling expectations.

\subsection{Least-squares learning and stochastic approximation \label{sec:LSL}}

As noted in Section \ref{sec:criteria}, an obvious approach to deliver on Criterion 3 is to model some form of learning. One approach that has a long tradition in the economics literature is \emph{least-squares learning}. See for example \citet{bray}, \citet{marcet-sargent}, \citet[][chapter 5]{sargent-bounded}, \citet{evans-honkapohja}, and \citet{evans-mcgough}. Least-squares learning is a special case of a more general set of \emph{stochastic approximation methods} \citep[e.g.][]{robbins-monro,ljung} which I also discuss below.\footnote{\citet{sargent-after-lucas} links stochastic approximation back to \citet{friedman-savage} and \citet{hotelling}.} Reinforcement learning, which I review in Section \ref{sec:RL}, is another stochastic approximation method.

\paragraph{Least-squares learning.} I review least-squares learning by means of a specific example: learning about equilibrium prices in the simple demand-supply system of  \citet{muth} which we already encountered in Section \ref{sec:muth}.\footnote{While I use somewhat different notation, this section follows closely the presentation in \citet{evans-honkapohja} who also start from Muth's example. Also see the presentation of \citet{bray} in chapter 5 of \citet{sargent-bounded} which starts directly from the reduced form \eqref{eq:reduced_form}, \citet{christiano-eichenbaum-johannsen}, and \citet{baley-veldkamp}.} Recall that production takes place with a one-period lag, that supply is an increasing function of the expected (log) price $p_t^e = \mathbb{E}_{t-1}[p_t]$ and of a supply shock $u_t$, and that equilibrium prices and quantities are determined by the intersection of demand and supply -- see \eqref{eq:muth}. As a result, equilibrium prices depend on price beliefs -- see \eqref{eq:reduced_form}.

Instead of assuming that producers have rational price expectations about next period's equilibrium price, we now assume that they form these expectations from past data. That is, producers form estimates $\widehat{p}_{t+1}^e = \widehat{\mathbb{E}}_{t}[p_{t+1}]$ from observations of past prices $p_1,...,p_{t}$ and supply shock realizations $u_1,...,u_{t}$ via least squares.

The simplest case is when the supply shock $u_t$ is serially \emph{un}correlated so that current and past supply conditions are uninformative about future prices. I cover this simple case first and then return to the more general case momentarily. When supply conditions $u_t$ are serially uncorrelated, producers do not take past $u_t$'s into account. Instead, they simply form the expectation $p_{t+1}^e$ by taking the sample average of past price observations
$$\widehat{p}_{t+1}^e = \frac{1}{t}\sum_{s=1}^{t} p_s.$$
To link this to least-squares estimation, consider a linear regression model with only an intercept and no independent variable: the sample average is then the least-squares estimate of the intercept parameter. Computation of this sample average can also be implemented recursively as follows:\footnote{The derivation is $\widehat{p}^e_{t+1} = \frac{1}{t}\sum_{s=1}^{t}p_s = \frac{1}{t}\left[p_t + \sum_{s=1}^{t-1}p_s\right] =\frac{1}{t}(p_t +(t-1) \widehat{p}^e_t) = \widehat{p}^e_t + \frac{1}{t}\left[p_t - \widehat{p}^e_t\right]$.}
\begin{equation}\label{eq:recursive_average}
\widehat{p}^e_{t+1} = \widehat{p}^e_{t} + \frac{1}{t}\left[p_t - \widehat{p}^e_{t} \right] 
\end{equation}
Notice how this update rule computes the new estimate $\widehat{p}^e_{t+1}$ from the previous estimate $\widehat{p}^e_{t}$ and the latest data $p_t$. This recursive implementation of the sample average is the special case of recursive least squares in a regression model with only an intercept.

More generally, when supply conditions $u_t$ are serially correlated, producers use past observations of $u_t$ to forecast prices. To this end, assume that $u_{t+1} = \rho u_{t} + \varepsilon_{t+1}$ where the correlation coefficient $\rho$ is unknown to producers and $\varepsilon_{t+1}$ is a mean-zero i.i.d. random variable. In this more general case, least-squares learning theory assumes that producers use a log-linear \emph{perceived law of motion}
$$
p_{t+1} = \theta_0 + \theta_1 u_t + \varepsilon_{t+1},
$$
where $\theta_0$ and $\theta_1$ are unknown parameters. That is, producers conjecture that the current supply shock $u_t$ forecasts next period's $u_{t+1}$ and hence next period's price $p_{t+1}$ but they do not know the strength of these relationships. To form price expectations $\widehat{p}_{t+1}^e = \widehat{\mathbb{E}}_{t}[p_{t+1}]$, producers then (i) estimate $(\theta_0,\theta_1)$ from observations of past prices $p_1,...,p_{t}$ and supply shock realizations $u_0,...,u_{t-1}$ via least squares, and (ii) use the time-$t$ estimates $(\widehat{\theta}_{0t},\widehat{\theta}_{1t})$ to compute $\widehat{p}_{t+1}^e = \widehat{\mathbb{E}}_{t}[p_{t+1}] = \widehat{\theta}_{0t} + \widehat{\theta}_{1t} u_t$.

Writing the perceived law of motion as $p_{t} = x_{t}^{\rm T}\theta + \varepsilon_t$ where $x_{t} = [1 \ u_{t-1}]^{\rm T}$ and $\theta = [\theta_0 \ \theta_1]^{\rm T}$, the time-$t$ least-squares estimate of the parameter vector $\theta$ is given by
$$\widehat{\theta}_{t} = \left[\sum_{s=1}^{t} x_s x_s^{\rm T} \right]^{-1}\left[\sum_{s=1}^{t} x_s p_s \right].$$
Analogously to \eqref{eq:recursive_average}, there is again a recursive implementation that computes the new estimate $\widehat{\theta}_{t+1}$ from the previous estimate $\widehat{\theta}_t$ and the latest data $(p_{t+1},u_t)$ -- see for example equation (2.9) in \citet{evans-honkapohja}. This recursive implementation is known as \emph{recursive least squares}. In summary, least-squares learning means that decision makers update parameter estimates for a perceived law of motion using recursive least squares.

In heterogeneous-agent models, least-squares learning can be applied to decision makers learning perceived laws of motion of equilibrium prices. \citet{jacobson} implements this approach in a housing model, where households learn about house prices using a simple variant of least-squares learning \citep[stochastic gradient learning as in][]{evans-honkapohja-SG} and uses it to study the U.S. housing boom of the 2000s. \citet{nakov-nuno} model overlapping generations of investors learning about equilibrium stock prices from experience \citep[as in][]{malmendier-nagel-depression-babies,malmendier-nagel-inflation}.

Related, there is a link between least-squares learning and Krusell-Smith methods (see Section \ref{sec:HA_forecasting}).\footnote{When discussing least-squares learning, \citet{sargent-palgrave} states: ``This is in effect what \citet{krusell-smith} do, though they do not connect their method to the learning literature." The connection is explicit in \citet{giusto} who studies least-squares learning about the aggregate capital stock (rather than prices) in the Krusell-Smith model.} In both approaches, decision makers use a (typically log-linear) perceived law of motion and estimate its coefficients via least squares. A difference is that least-squares learning implements this coefficient estimation recursively so that solving for equilibrium and least-squares estimation (belief updating) are done ``in one sweep." This also means that coefficient estimates vary over time (which they do not in Krusell-Smith) and makes least-squares learning more likely to generate interesting phenomena like booms and busts \citep{jacobson}.

\paragraph{Stochastic approximation.} The update rule \eqref{eq:recursive_average} for the sample-average case and the recursive least-squares update rule for $\widehat{\theta}_t$ for the more general case take a common form that also arises in a number of other problems (for example, in the next subsection). Using the computer science notation to let $\leftarrow$ denote an assignment statement, \citet[][chapter 2.4]{sutton-barto} write the general form for this type of update rule as
\begin{equation}\label{eq:SA_update}
\mbox{NewEstimate} \leftarrow \mbox{OldEstimate} + \mbox{StepSize}\left[\mbox{Target} -  \mbox{OldEstimate}\right].
\end{equation}
The ``Target" is the target value that the algorithm attempts to drive the estimate to, i.e. it indicates the direction in which the estimate is updated, while the step size controls how large this update is.
%\footnote{$$\mbox{NewEstimate} - \mbox{Target} = (1 - \mbox{StepSize})\left[ \mbox{OldEstimate}- \mbox{Target}\right]$$} 
For example, in the sample-average case \eqref{eq:recursive_average}, the target is the time-$t$ price $p_t$ meaning that the new estimate $\widehat{p}_{t+1}^e$ is updated in the direction of $p_t$, and the step size is $1/t$ so that updates become smaller and smaller over time. This type of iterative update method is known as a \emph{stochastic approximation method} \citep[e.g.][]{robbins-monro,ljung}. See \citet[][ch.6]{zhao} for a good textbook discussion.

Casting least-squares learning as a stochastic approximation method is useful because convergence results exist for the latter. These results answer questions about the convergence of the least-squares coefficient estimates $\widehat{\theta}_t$ as $t \rightarrow \infty$ (as data becomes more and more plentiful). To this end, denote the step size in the update rule \eqref{eq:SA_update} by $\alpha_t$, for example $\alpha_t = 1/t$ in the sample-average case \eqref{eq:recursive_average}.\footnote{Considering more general step sizes links the update rule \eqref{eq:recursive_average} to adaptive expectations. For a general step size $\alpha_t$ the update rule becomes $\widehat{p}^e_{t+1} = \widehat{p}^e_{t} + \alpha_t\left[p_t - \widehat{p}^e_{t} \right]$. In the special case of a constant step size $\alpha_t=\alpha$, this becomes the adaptive expectations rule $\widehat{p}^e_{t+1} = \widehat{p}^e_{t} + \alpha\left[p_t - \widehat{p}^e_{t} \right]$ which implies $\widehat{p}^e_{t+1}  = (1-\alpha)^{t+1}\widehat{p}^e_{0}+ \alpha\sum^{t}_{j=0}(1-\alpha)^jp_{t-j}$, a backward-looking weighted average of past prices with exponentially declining weights.} One key result in stochastic approximation theory states that, as $t \rightarrow \infty$, the estimate being updated according to \eqref{eq:SA_update} converges with probability $1$ under a number of conditions including the following conditions on the step size:\footnote{As \citet{sutton-barto} explain: \emph{``The first condition is required to guarantee that the steps are large enough to eventually overcome any initial conditions or random fluctuations. The second condition guarantees that eventually the steps become small enough to assure convergence."}}
\begin{equation}\label{eq:SA}
\sum_{t=0}^\infty \alpha_t = \infty \qquad \mbox{and} \qquad \sum_{t=0}^\infty \alpha_t^2 < \infty.
\end{equation}
Both convergence conditions are met for the sample-average case in which $\alpha_t = 1/t$. More specifically, under these conditions, as $t \rightarrow \infty$, the stochastic trajectory of the estimate satisfying the update rule \eqref{eq:SA_update} becomes closer and closer to the deterministic trajectory of a particular ordinary differential equation (ODE) derived from this update rule. If this ODE has a locally stable steady state, the estimate converges to a constant that equals this steady state value. See \citet{ljung} as well as the discussions in \citet{marcet-sargent}, \citet{woodford-sunspots}, \citet{evans-honkapohja}, and \citet{sargent-conquest}.

Using stochastic approximation theory, the literature has shown that, under certain conditions, the equilibrium with least-squares learning converges to the rational expectations equilibrium \citep{marcet-sargent,woodford-sunspots}. \citet{evans-honkapohja} dubbed these conditions ``expectational (E) stability." For example, in the simplest case of the Muth model without serial correlation, it is intuitive that the backward-looking sample average (recursive least-squares expectation) $\widehat{\mathbb{E}}_{t}[p_{t+1}] = \frac{1}{t}\sum_{s=1}^{t} p_s$ converges to the rational expectation $\mathbb{E}_{t}[p_{t+1}]$ as $t\rightarrow \infty$. Similar results hold when there is serial correlation or for more general models, though these results require relatively strong assumptions and can be fragile.

As discussed in Section \ref{sec:criteria}, to actually simplify computations in economies with heterogeneity (Criterion 1), the perceived law of motion of equilibrium prices would need to be ``restricted"  to \emph{not} start from rational expectations as a special case. This implies that the least-squares learning process cannot converge to the rational expectations equilibrium. But it may instead converge to a ``restricted perceptions equilibrium" (Appendix \ref{app:equilibria}) which may therefore be a relevant equilibrium concept in heterogeneous-agent models.

\paragraph{Slow or non-convergent learning.} An important caveat is that stochastic approximation convergence theorems hold only asymptotically as $t\rightarrow \infty$ and that learning may be slow in practice. See \citet{christiano-eichenbaum-johannsen} as well as \citet{sargent-bounded} who voices reservations about adaptive algorithms as theories of real-time dynamics but states that he nevertheless ``like[s them] as devices for selecting equilibria." If the environment is sufficiently complex and non-stationary, it is also possible that learning never converges \citep[e.g.][]{bouchaud-farmer,garnierbrun-benzaquen-bouchaud}.

\paragraph{Learning as process.} In contrast to rational expectations, learning models explicitly specify the \emph{process} by which individuals form expectations. This approach echoes \citeauthor{simon-rationality}'s (\citeyear{simon-rationality}) distinction between ``procedural" and ``substantive" rationality. In a similar vein, \citet{simon-decide} writes: \emph{``The theory of the business cycle is another important candidate area, for a procedural theory of the forming of expectations and the making of decisions. [...] One direction of progress is to erect theories that postulate [...] how expectations for the future are in fact formed by economic actors, and how those expectations enter into the calculations of actions. A realistic procedural theory would almost certainly have to include learning mechanisms."} As an aside, \citet{simon-decide} also features a fascinating discussion of artificial intelligence from the perspective of its time.

\subsection{Reinforcement Learning \label{sec:RL}}
Reinforcement learning (RL) is learning value functions of incompletely-known Markov decision processes. Decision makers do not know the exact environment they are operating in (which may be extremely complex) and instead learn optimal policies from experience. RL is at the core of some impressive advances in artificial intelligence, e.g. learning to play Go and Atari games better than humans \citep{deepmind-atari,deepmind-go,deepmind-go-zero} and psychologists have argued that RL is responsible for part of human and animal learning \citep[e.g.][]{niv,glimcher,caplin-dean,neuroeconomics-book,gershman-daw}.\footnote{RL ideas have also been applied in game theory \citep[e.g.][]{roth-erev-95,erev-roth,fudenberg-levine} but using a different formulation that does not work with value functions and instead directly reinforces the ``propensities" of choosing strategies. Also see \citet{barberis-jin}, \citet{chen-etal-RL-monetary}, \citet{duarte-duarte-silva,duarte-fonseca-goodman-parker}, \citet{barrera-desilva}, and the references in \citet{FV-nuno-perla} for applications in finance and macroeconomics.}

RL ideas seem to me a promising direction for developing alternative approaches to rational expectations about equilibrium prices in heterogeneous-agent models: a natural assumption is that agents do not know the true stochastic process for equilibrium prices (which is extremely complex, with prices being driven by an underlying Markov process for an entire cross-sectional distribution -- see section \ref{sec:not_markov}) and instead learn about prices in some way.\footnote{A number of papers have applied RL to solving heterogeneous-agent models \emph{without} aggregate risk. See for example \citet{xu-etal-RL} and \citet{lauriere-etal-RL}. As should hopefully be clear, the present paper is instead about models \emph{with} aggregate risk.} RL is linked to least-squares learning because both are stochastic approximation methods (see the previous subsection).

\begin{figure}[htp]
	\begin{center}
	\includegraphics[width=.45\textwidth]{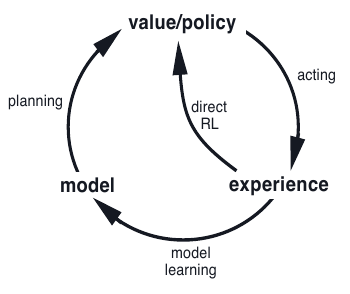}
	\end{center}
\vspace{-3mm}
	\begin{minipage}{\textwidth}
		\caption{Model-free vs. model-based reinforcement learning \citep[][ch.8.2]{sutton-barto}. \label{fig:RL}} 
	\end{minipage}
\end{figure}

\paragraph{Reinforcement learning in a nutshell.}
I briefly summarize the basics of RL following the brilliant introductory treatments by \citet{sutton-barto} and \citet{zhao}.\footnote{Also see \citet{murphy}, \citet{silver-RL-course} for an online course with video recordings, and \citet{charpentier-elie-remlinger}.} RL is concerned with Markov decision processes (MDPs) of the type that are very familiar to economists: an agent maximizes the expected present-discounted value of period rewards
$$\mathbb{E}_0 \sum_{t=0}^\infty \beta^t R(x_t,c_t), $$
where $x_t \in \mathcal{X}$ is a state, $c_t \in \mathcal{C}$ is a control (action), $R$ is a reward function, and $0<\beta<1$ is a discount factor. The evolution of the state $x_t$ is stochastic and depends on the control $c_t$ according to the transition probabilities $x_{t+1} \sim P(\cdot|x_{t},c_{t})$. My notation differs from the standard computer science notation \citep{sutton-barto,zhao}  to make it closer to the notation commonly used in economics. As is standard, the associated Bellman equation for the value function is
$$v(x) = \max_c \ R(x,c) + \beta \int_{x'} v(x') P(x'|x,c) .$$
The difference to standard dynamic programming is that RL is concerned with the case where the agent \emph{does not know the model}, in particular the transition probabilities $P$. That is, RL is the \emph{optimal control of incompletely-known Markov decision processes}. Somewhat more precisely, agents only see the reward $R_t$ they receive after taking action $c_t$ (as well as the current state $x_t$ and successor state $x_{t+1}$). Over time, agents learn to take actions that maximize rewards. Rewards are thus like ``doggy treats" that reinforce good behavior and this connection to animal psychology is how reinforcement learning got its name. A noteworthy feature of standard RL is that it is ``model-free" in that agents learn directly from experience rather than trying to learn the underlying model first (model-based RL). Figure \ref{fig:RL} from \citet{sutton-barto} illustrates the distinction between model-free (or direct) RL and model-based RL.

In mathematical terms, the key idea of RL is to learn value functions from experience (observations of states, actions, and rewards) by means of a stochastic approximation algorithm of the type we encountered in the previous subsection \citep{jaakkola-jordan-singh,tsitsiklis}. This links RL to least-squares learning and other instances of stochastic approximation algorithms.

To understand how this works, consider first the problem of learning the value function for a fixed deterministic policy $\pi$ which prescribes a control in feedback form $c = \pi(x)$. Fixing the policy function means that there is no decision and hence this is what the literature calls a Markov reward process (MRP). The corresponding value function is
$$v_\pi(x) = R(x,\pi(x)) + \beta \mathbb{E}_{\pi}[v_\pi(x_{t+1})|x_t = x].$$
The standard method for approximating this value function is via \emph{temporal-difference learning} (TD learning). A sketch of the algorithm is as follows:
\begin{itemize}
\item Input: a policy $\pi(x)$, a step-size parameter $\alpha>0$, and a number of time steps $T$ (a large number, say $T=1000$).
\item Initialize $V(x)$ with some initial guess for all $x\in \mathcal{X}$.
\item Then, starting from a randomly drawn initial $x_0$, for each time step $t=0,1,...,T$:
\begin{enumerate}
\item given the current state $x_t$ and policy $\pi$, compute the next state $x_{t+1}$ using the transition probabilities $P(\cdot|x_t,\pi(x_t))$
\item value prediction: update the guess for the value function according to
\begin{equation}\label{eq:TD0}
V(x_t) \leftarrow V(x_t) + \alpha[\underbrace{R_{t} + \beta V(x_{t+1})}_{\mbox{\footnotesize Target}} - V(x_t)].
\end{equation}
\end{enumerate}
\end{itemize}
This algorithm returns a value function $V(x)$ for all $x \in \mathcal{X}$. The recursive updating step \eqref{eq:TD0} takes the form of the general stochastic approximation updating rule \eqref{eq:SA_update} and consequently, under certain conditions, $V(x)$ converges to $v_\pi(x)$ as experience $T\rightarrow \infty$ \citep{jaakkola-jordan-singh,tsitsiklis,sutton-barto,zhao}. 

The full RL problem is to also solve for the optimal policy $\pi$. Because the agent does not know the transition probabilities $P$, it does not suffice to learn the value function $v_\pi(x)$. Instead, the agent must learn a so-called \emph{action-value function} or \emph{Q function} which equals the value of taking action $c$ in state $x$. This action-value function is defined for state-action pairs $(x,c)$ as $Q(x,c):= R(x,c) + \beta \int_{x'} v(x') P(x'|x,c)$ and satisfies the Bellman equation
$$Q(x,c)= R(x,c) + \beta \int_{x'} \max_{c'} Q(x',c') P(x'|x,c).$$
Given $Q$, the optimal policy can then be computed as $\pi(x) = \arg \max_c Q(x,c)$. The agent learns this action-value function using variants of the TD algorithm described above. Two common algorithms are \emph{Sarsa} (which is the most straightforward conceptually) and \emph{Q-learning}.\footnote{The name \emph{Sarsa} comes from the standard computer science notation which denotes states by $S_t$, actions by $A_t$ and rewards by $R_t$ and that agents update action-value functions from the sequence $(S_t,A_t,R_t,S_{t+1},A_{t+1})$.} RL algorithms for finding optimal policies also involve some variant of ``exploration", i.e. trying out new policies, as opposed to only using policies that maximize the current guess of the action-value function (``exploitation"). For example, so-called $\varepsilon$-greedy policies choose actions that are suboptimal given the current action-value function with small probability $\varepsilon$.

\paragraph{Big World Hypothesis.} Despite exponentially growing computation, applications of RL to real-world learning problems have remained limited. \citet{javed-sutton} argue that in many learning problems RL agents operate in ``big worlds", i.e. environments that are orders of magnitude more complex than the agent's information processing capacity. They argue that algorithms should reflect this reality of ``small bounded agents learning in large unbounded environments" and propose an approach to algorithm design reminiscent of ideas in the bounded rationality and non-rational expectations literatures in economics.

Heterogeneous-agent economies (and indeed real-world macroeonomies) seem to me good examples of big worlds and so some of these ideas may be applicable to the challenge posed in this paper. The introduction of \citet{javed-sutton} illustrates the parallels: \emph{``The big world hypothesis says that in many decision-making problems the agent is orders of magnitude smaller than the environment. It can neither fully perceive the state of the world nor can it represent the value or optimal action for every state. Instead, it must learn to make sound decisions using its limited understanding of the environment. The key research challenge for achieving goals in big worlds is to come up with solution methods that efficiently utilize the limited resources of the agent."}

%Other contributions have pursued similar ideas. The RL agent of \citet{dong-vanroy-zhou} can operate in complex non-Markovian environments which may be relevant for heterogeneous-agent economies with non-Markovian equilibrium prices (Section \ref{sec:not_markov}). The agent of \citet{vanroy-etal-constrained} learns subject to an entropy-based information processing constrained reminiscent of (though different from) the rational inattention literature.

\paragraph{Reinforcement learning about equilibrium prices?} In ongoing work with coauthors, I am exploring  whether RL ideas can be adapted to the problem of learning about equilibrium prices in heterogeneous-agent models with aggregate risk. One idea is that agents do not know the correct transition probabilities of equilibrium prices (which are extremely complex) and instead learn about the price process from experience using TD learning. The hope is that this type of approach may be able to satisfy Criteria 1 to 3 above. While TD learning about equilibrium prices would not feature an explicit model of price beliefs, the implicit beliefs would still be consistent with model reality (Criterion 3a) in that mistakes that are costly in terms of values would be avoided. Along the lines discussed in Section \ref{sec:not_markov}, adapting ideas from the theory of RL in non-Markovian environments or POMDPs could be a fruitful avenue. 

An interesting question is whether delivering on Criterion 2 (consistency with empirical evidence) will require modifying standard RL ideas. While RL will converge to an equilibrium different from rational expectations, a natural conjecture is that the resulting RL expectations will be ``close to'' rational. This suggests that modifications of standard RL ideas may be needed to take on board empirical evidence on belief formation.

Another interesting question is whether standard model-free RL can satisfy Criterion 3 or whether a more model-based approach is needed. An advantage of model-free RL over typical adaptive learning methods in economics is that it avoids specifying a perceived law of motion for prices, which necessarily involves some arbitrariness. However, a completely model-free approach, in which decision makers lack any structural understanding of the world, may be too simplistic \citep[e.g.][]{enke-cognitive-turn}. For both model-free and model-based RL methods, learning may also be slow in practice -- see the discussion on convergence speed above. Since RL and least-squares learning are both special cases of stochastic approximation algorithms, this concern applies equally to both.

RL is a machine learning method so it is natural to ask how such an approach relates to recent work applying deep learning methods to solve heterogeneous agent models with aggregate risk \citep[e.g.][]{han-yang-e,azinovic-gaegauf-scheidegger,gu-lauriere-merkel-payne,gopalakrishna-gu-payne}.\footnote{With the exception of \citet{han-yang-e} these papers do not use RL ideas and instead use deep neural networks as a powerful function approximation method. Yet another approach is to use a deep neural network to approximate a perceived law of motion for distributional moments \citep[e.g.][]{maliar-maliar-winant,FV-hurtado-nuno,FV-marbet-nuno-rachedi}, i.e. a non-linear version of \citet{krusell-smith}.} These papers typically use deep neural networks to approximate the Master equation, i.e. to approximate value or policy functions on a state space that includes the high-dimensional cross-sectional distribution. Such work uses deep learning to ``tame the curse of dimensionality" \citep{FV-nuno-perla}, whereas RL about equilibrium prices would instead \emph{sidestep} this curse by modeling decision makers as solving lower-dimensional problems.

\subsection{Heuristics and Simple Models\label{sec:simple}}
Like the real world, heterogeneous-agent economies are complex environments that generate complex stochastic processes for equilibrium prices. A natural conjecture is that households and firms operating in this type of complex, stochastic environment employ heuristics and simplified models when making decisions under uncertainty. This idea has a long tradition, see for example \citet{tversky-kahnemann-science} who argue that \emph{``people rely on a limited number of heuristic principles which reduce the complex tasks of assessing probabilities and predicting values to simpler judgmental operations"} and \citet{gigerenzer-todd}. 

In the macroeconomics literature, a number of authors have pursued the related idea that decision makers use simplified, restricted models to forecast economic variables. See for example \citet{sargent-RPE}, \citet{brock-hommes}, \citet{degrauwe}, \citet{fuster-hebert-laibson-JEL,fuster-hebert-laibson-MA}, \citet{molavi}, \citet{hajdini}, and \citet{flynn-sastry}. Many (but not all) of these authors impose some consistency between subjective beliefs and model reality as in a restricted perceptions equilibrium (Appendix \ref{app:equilibria}). This approach -- assuming that decision makers form beliefs using ``simple models" and finding a restricted perceptions equilibrium -- could be applied to price beliefs in heterogeneous-agent models and may be able to satisfy my three criteria. First, the idea that decision makers use simple models and heuristics goes well with Criterion 1 (computational tractability), in particular the point that models of non-rational expectations should not require decision makers to compute the special case with rational expectations. Second, to make such simple models consistent with key empirical evidence (Criterion 2), one could use popular behavioral biases like diagnostic expectations or cognitive discounting to ``distort" the simplified forecasts (not the rational ones). Third, the restricted perceptions equilibrium would deliver Criterion 3 (endogeneity of beliefs to model reality).

The relative sophistication of such simplified forecasting models may differ across decision makers. In particular, financial market participants may be more sophisticated than ordinary households but, over time, the latter may incorporate some of the sophistication of the former \citep{caballero-simsek}. However, even the most sophisticated financial players do not forecast prices by forecasting entire (infinite-dimensional) cross-sectional distributions.

Finally, a recent literature tries to better understand the cognitive foundations of economic decision making in the face of risk and complexity, i.e. to understand ``how people think." See \citet{enke-cognitive-turn} for a review and \citet{oprea} for an example. \citet{enke-cognitive-turn} points to evidence that some decision makers fully neglect certain aspects of complex environments (``incomplete representations"), in particular indirect or general equilibrium effects (``system neglect"). Incorporating ideas from this literature could be another way forward.

%\citet{caballero-caravello-simsek}

\section{Conclusion \label{sec:conclusion}}
This paper has posed a challenge: to develop alternatives to the assumption of rational expectations about equilibrium prices in heterogeneous agent economies. To this end, I outlined three criteria that alternative approaches should fulfil:  (1) computational tractability, (2) consistency with empirical evidence, and (3) endogeneity of beliefs to model reality (Lucas critique). I then discussed some promising directions, including temporary equilibrium approaches, incorporating survey expectations, least-squares learning, and reinforcement learning.

Developing alternatives to rational expectations holds two main promises. First, to bring the empirical and theoretical discipline of heterogeneous-agent macroeconomics to the study of economic booms and busts and other important questions in which aggregate non-linearities are key. Second, to make these models more realistic, more consistent with empirical evidence on expectations formation, and more likely to generate booms and busts and other interesting phenomena in the first place. While there is often a trade-off between realism and (computational) simplicity, this could be a rare case in which the two go hand in hand.

Macroeconomists specializing in heterogeneous-agent models and those working on the theory and empirics of belief formation should work together and capitalize on the resulting gains from trade. Work combining ideas from these two literatures is starting to emerge. See for example the work on temporary equilibrium with survey expectations discussed above as well as \citet{guerreiro}, \citet{andre-etal-narratives}, \citet{baley-turen}, and \citet{cai}. Similarly, \citet{broer-etal-KS} call for work that considers ``more general, dynamic information-choice strategies that can simultaneously match the rich micro-heterogeneity in expectations and explore the subsequent macroeconomic implications."\footnote{\citet{broer-etal-heterogeneity} take some steps in this direction but in a way that, counter to my first criterion, makes decision makers' price forecasting problem more complicated than in the Master equation: the cross-sectional wealth distribution remains a state variable in decision makers' Bellman equations but, additionally, so is a cross-sectional distribution of beliefs over such wealth distributions.} I agree. However, it is also important that researchers not only consider deviations from rational expectations in linear or linearized models (as they often do).

In a similar vein, we need to develop efficient \emph{global} solution methods for heterogeneous agent models. By focussing on linear or other local solution methods, the heterogeneous-agent literature risks repeating the mistakes of the pre-financial crisis representative-agent literature. See for example \citet{blanchard-danger} who writes: \emph{``The reason for this assumption, called linearity, was technical: models with nonlinearities [...] were difficult, if not impossible, to solve under rational expectations. Thinking about macroeconomics was largely shaped by those assumptions. We in the field did think of the economy as roughly linear, constantly subject to different shocks, constantly fluctuating, but naturally returning to its steady state over time. [...] The problem is that we came to believe that this was indeed the way the world worked."} A weaker form of Blanchard's criticism applies more generally to methods that solve models only locally around a steady state and which may therefore miss important global dynamics -- see Section \ref{sec:global}. Such local methods remind me a bit of the old joke about the drunk who is looking under a lamppost for a key that he has lost on the other side of the street because ``this is where the light is."

The theoretical expectations literature should also allocate more time to modeling expectations about endogenous variables like equilibrium prices \citep[as in][for example]{bastianello-fontanier}. Finally, the literature should focus less on developing models of non-rational expectations that require decision makers to compute the special case of rational expectations. In the context of this paper, this defeats the very purpose of departing from rational expectations.

\bigskip
\bigskip
\bigskip

\appendix

\setstretch{1.25}
\renewcommand{\thefigure}{A\arabic{figure}}
\setcounter{figure}{0}
\renewcommand{\thetable}{A\arabic{table}}
\setcounter{table}{0}
{\noindent\LARGE \textbf{Appendix}}

\section{A simple prototype for models requiring global solution methods}\label{app:global}

To complement the discussion in Section \ref{sec:global}, this appendix spells out a simple example of a stochastic process that serves as a useful prototype for more complicated macroeconomic models featuring phenomena like state spaces with crisis regions or bimodal ergodic distributions for the aggregate economy. Such models need to be solved using global solution methods because local perturbation methods fail to capture their behavior.

This prototype is the scalar ``double-well" or ``bistable" diffusion process
\begin{equation}\label{eq:bistable}
dX_t = -V'(X_t)dt + \sigma dW_t,
\end{equation}
where $\sigma$ parameterizes uncertainty, $W_t$ is a standard Brownian motion, and where $V(x)$ is a ``double-well potential" function, that is a function that has two minima at $a$ and $c$ and a local maximum at $b$ in between. For example, $V(x) = \frac{1}{4}x^4 - \frac{1}{2}x^2$ which has minima at $-1$ and $1$ and a maximum at $0$. Equivalently, $-V'(x)$ is $S$-shaped and intersects zero three times. Good treatments are in  \citet[][ch.14]{gardiner} and \citet[][ch.7]{pavliotis}. An analogous discrete-time formulation is $X_{t+1} = S(X_t) + \sigma \varepsilon_t$ where $\sigma$ parameterizes uncertainty, $\varepsilon_t$ is a random variable, and $S(x)$ is an S-shaped function that intersects the 45-degree line three times. See \citet{azariadis-stachurski} and \citet{bouchaud-morelli-etal} for discrete-time models with such dynamics.

Without uncertainty $\sigma=0$ the double-well process \eqref{eq:bistable} has three steady states, two of them stable and one unstable. The two stable steady states are at $a$ and $c$ and the unstable steady state is at $b$, i.e. the two minima and the maximum of $V$.

In contrast, with uncertainty $\sigma>0$, this process has a bimodal stationary distribution\footnote{From the stationary Kolmogorov Forward equation $0 = (V'(x)f(x))' + \frac{\sigma^2}{2} f''(x)$ for all $x \in \mathbb{R}$.}
$$f(x) \propto \exp(-2 V(x)/\sigma^2)$$
in which the two modes (the two maxima of $f$) are the two minima of $V$, i.e. points $a$ and $c$. See for example, Figure 14.1 in \citet{gardiner} or Figure 7.2 in \citet{pavliotis}. The system is thus most likely to be found at $a$ and $c$. 

Trajectories of $X_t$ spend most time oscillating around the two stable steady states $a$ and $c$ while occasionally hopping between them -- see, for example, Figure 7.2 in \citet{pavliotis}. Starting from the high steady state, ``the economy" stays close to it most of the time but may be thrown into a ``crisis" (low steady state) and get stuck there for a while before ultimately recovering.  

Perturbation methods around $\sigma=0$ completely miss this type of behavior. See \citet[][ch.7.2.4]{gardiner} for a good discussion. Intuitively, in the words of \citet{blanchard-danger} cited in the conclusion, a perturbation method would think of the system \emph{``as roughly linear, constantly subject to different shocks, constantly fluctuating, but naturally returning to its steady state over time,"} but this does not reflect the system's actual dynamics.

\section{Various concepts of non-rational expectations equilibrium \label{app:equilibria}}
The economics literature has proposed various concepts of ``non-rational expectations equilibrium" or ``misspecification equilibrium." This appendix briefly summarizes a number of these and puts them in relation to each other. As already stated in Section \ref{sec:TE}, using the acronyms defined there or below, the relation between the various equilibrium concepts is summarized in \eqref{eq:equilibria} which I restate here for the reader's convenience:
\begin{equation*}
\{REE,RPE,CEE\} \subset SCE \subset IREE \subset TE,
\end{equation*}
where $\subset$ means ``is a special case of."

\paragraph{Self-confirming equilibrium (SCE).} See for example \citet{sargent-conquest},  \citet{cho-sargent}, and \citet{fudenberg-levine} in the context of games. In a self-confirming equilibrium, actual equilibrium outcomes are statistically consistent with decision makers' beliefs, i.e. these beliefs are not disappointed. Self-confirming equilibria can be limiting outcomes of adaptive learning processes of the type described in Section \ref{sec:LSL}. A rational expectations equilibrium is a self-confirming equilibrium, but not vice versa. Specifically, beliefs may be incorrect for events that are infrequently observed (e.g. events off the equilibrium parts).

\paragraph{Restricted perceptions equilibrium (RPE).} See for example \citet{sargent-RPE} and \citet{branch}. An RPE is a SCE in which decision makers use restricted forecasting models that do not nest rational expectations, i.e. mis-specified perceived laws of motion. As a result, a RPE is never a REE. Learning processes that update restricted forecasting models may settle down to a RPE (and never a REE). As discussed in Section \ref{sec:criteria}, to actually simplify computations in heterogeneous-agent economies (Criterion 1), candidate alternative approaches should \emph{not} start from rational expectations as a special case and so RPE is a natural equilibrium concept. See \citet{baley-turen} for a recent application of RPE in a model of forecasters' heterogeneous inflation expectations.
 
\paragraph{Consistent expectations equilibrium (CEE).} See \citet{hommes-sorger}. In a CEE decision makers use autoregressive forecasting models such as an $AR(p)$ process, with a consistency condition for the autocorrelations as for SCE. A CEE is thus a special case of a SCE. In principle, a CEE could coincide with the REE if the RE solution happened to be exactly of the same autoregressive structure as the assumed forecasting model. But RE solutions typically do not take this form (and certainly RE equilibrium prices in  heterogeneous-agent models do not) so the CEE forecasting model is typically mis-specified. Thus a CEE is typically a special case of a RPE \citep{branch} so that \eqref{eq:equilibria} could also write $CEE \subset RPE$.

\paragraph{Another equilibrium concept not included in \eqref{eq:equilibria}.} \emph{Oblivious equilibrium} is an equilibrium concept for stochastic dynamic games like dynamic industry models of imperfect competition \citep{weintraub-benkard-vanroy-NEURIPS,weintraub-benkard-vanroy,weintraub-benkard-vanroy-computation}. These games suffer from a curse of dimensionality similar to the one in heterogeneous-agent models: in a Markov Perfect Equilibrium, the distribution of other players' states is a state variable in player's dynamic programming problem. Oblivious equilibrium restricts players' policies to be ``oblivious" to changes in the distribution of other players' states and instead restricts these policies to be functions of the constant long-run average of this distribution.\footnote{The standard formulation only considers models without aggregate shocks and so is not relevant for our purposes. \citet{weintraub-benkard-vanroy-computation} extend the approach to aggregate shocks by restricting strategies to depend only on a low-dimensional vector that is a function of the history of the aggregate shock (e.g. a truncation of the history).}

\section{The \citet{morgenstern} critique of perfect foresight in general equilibrium 
\label{sec:morgenstern}}

In an article written ninety (!) years ago, \citet{morgenstern} criticized the assumption of perfect foresight in general equilibrium. This article is relevant because perfect foresight is the precursor  to -- or, indeed, special case of -- rational expectations in models without uncertainty. The article is in German and is entitled ``Vollkommene Voraussicht und wirtschaftliches Gleichgewicht" which translates to ``Perfect Foresight and General Equilibrium." To make it more accessible to English readers, I am including a translation of some key passages. The same passages in the original German are further below.

\paragraph{English Translation.} I translated the passages using ChatGPT and then edited for clarity.

\citet{morgenstern}, p.337: \emph{``The pride of theoretical economics is the theory of general economic equilibrium, which has been developed in various forms. [...] [The following] remarks are intended to draw attention to a problem of equilibrium theory -- and thus of every variety of theoretical economics [...]. This concerns the assumption of what is here used synonymously as `full foresight' or `perfect foresight,' which is allegedly one of the preconditions of equilibrium."}

p.342: \emph{``It is `economic' things and events that are to be foreseen. On the admissible assumption that it is exactly known what is meant by this (for example, prices, production yields, etc.), one finds that, owing to the interdependence of all economic processes and conditions with one another, and of these with all other facts, no matter how small a segment of events could be specified whose foresight would not at the same time entail foresight of the entire remainder. [..] The most important and ultimately decisive elements of this kind are the individual acts of behavior from which the complex magnitudes arise."}

\emph{``The forward-looking individual must therefore know not only exactly the influence of his own actions on prices, but also that of all other individuals, and that of his own future behavior on the behavior of others -- especially those who are personally relevant to him. The circle of these relevant individuals is extraordinarily large, since all indirect effects must also be foreseen exactly. [Perfect foresight] also leads to the result that individuals must possess complete insight into  theoretical economics -- which is only to be supplied by equilibrium theory itself; for how else should they be able to foresee long-range effects?"}

\emph{``The improbably high demands placed on the intellectual capacity of economic agents show at the same time that the equilibrium systems do not encompass ordinary human beings, but at least demigods who are exactly alike among themselves, if indeed the requirement of full foresight is to be met. This, then, is of no use whatsoever. If “full” or “perfect” foresight -- in the strictly definable sense evidently intended by the economic authors, namely of unlimited foresight -- is to be taken as a premise of equilibrium theories, then it is a completely nonsensical assumption."}

p.345: \emph{``The necessity that, with perfect foresight, each individual must grasp all economic interrelations -- that is, must master economic theory -- leads to a scientifically and logically curious fact. If perfect foresight were an indispensable condition for formulating general equilibrium theory, it would result in the further paradox that science is already presupposed in the very object it is meant to investigate. [...] This logical-scientific issue is most clearly illustrated by comparison with the natural sciences. In physics or chemistry, there is absolutely no assumption that physical or chemical laws are known by the very objects these sciences aim to explain -- for example, that atoms need to make assumptions about the behavior and states of other atoms."}

\paragraph{Original German.} In the original German, the same passages are as follows:

\citet{morgenstern}, S.337: \emph{``Den Stolz der theoretischen Ökonomie bildet die Theorie des allgemeinen wirtschaftlichen Gleichgewichtes, die in verschiedenen Formen entwickelt worden ist. [Die nachfolgenden] Ausführungen bezwecken auf ein Problem der Gleichgewichtstheorie -- und damit jeder Abart von theoretischer Ökonomie -- hinzuweisen [...]. Es handelt sich um die Annahme der (hier synonym gebrauchten) `vollen Voraussicht' oder `vollkommenen Voraussicht', die angeblich eine der Vorbedingungen des Gleichgewichtes ist."}

S.342: \emph{``Vorausgesehen werden sollen `wirtschaftliche' Dinge und Ereignisse. Unter der zulässigen Annahme, es sei genau bekannt, was darunter zu verstehen ist (z.B. Preise, Produktions­erträge usw.), findet man, daß infolge der Interdependenz aller wirtschaftlichen Prozesse und Gegebenheiten untereinander und dieser mit allen anderen Tatsachen kein noch so kleiner Ausschnitt aus dem Geschehen angegeben werden könnte, dessen Voraussicht nicht zugleich die Voraussicht des gesamten Restes bedeutete. [...] Die wichtigsten und letztlich entscheidenden Elemente dieser Art sind die individuellen Verhaltensakte aus denen die komplexen Größen hervorgehen."}

\emph{``Das vorausschauende Individuum muß also nicht nur genau den Einfluß seines eigenen Handelns auf die Preise kennen, sondern auch den aller anderen Individuen und den seines eigenen zukünftigen Verhaltens auf das der anderen, namentlich der für ihn persönlich relevanten. Der Kreis dieser relevanten Individuen ist außerordentlich groß, da doch auch alle indirekten Wirkungen genau mitvorausgesehen werden müssen. [Die `vollkommene Voraussicht'] führt übrigens dazu, daß die Individuen auch eine vollständige Einsicht in die -- erst durch die Gleichgewichtstheorie zu liefernde -- theoretische Ökonomie haben müssen, denn wie anders sollten sie sonst die Fernwirkungen voraussehen können?"}

\emph{``Die unwahrscheinlich hohen Ansprüche, die an die intellektuelle Leistungsfähigkeit der Wirtschaftssubjekte gestellt werden, beweisen zugleich, daß in den Gleichgewichtssystemen keine gewöhnlichen Menschen erfaßt werden, sondern mindestens untereinander genau gleiche Halbgötter, falls eben die Forderung voller Voraussicht erfüllt sein soll. Damit ist also nichts anzufangen. Wenn `volle' oder `vollkommene' Voraussicht im streng angebbaren und von den ökonomischen Autoren offenbar gemeinten Sinne einer unbeschränkten Voraussicht den Gleichgewichtstheorien zugrunde gelegt werden soll, so handelt es sich um eine völlig sinnlose Annahme."}

S.345: \emph{``Die Notwendigkeit, dass jedes Individuum bei völliger Voraussicht alle wirtschaftlichen Zusammenhänge überschauen, also die theoretische Ökonomie beherrsehen muss, führt zu einer wissenschaftslogisch merkwürdigen Tatsache. Ware völlige Voraussicht eine unerlässliche Bedingung für die Aufstellung der Gleichgewichtstheorie, so ergäbe sich das weitere Paradox, dass die Wissenschaft bei dem Objekt, das sie erst erforschen soll, schon vorausgesetzt wird [...] Die wissenschaftslogische Situation ist am klarsten gegenüber den Naturwissenschaften darzutun. Bei der Physik oder Chemie wird in gar keiner Weise die Präexistenz physikalischer oder chemischer Lehrsätze bei den von diesen Wissenschaften zu erklärenden Objekten -- z.B. den Atomen und Elementen -- vorausgesetzt, derart, daß die Atome Annahmen über das Verhalten und die Zustände der anderen Atome machen müßten."}

%
%%\citet{markiewicz-pick}
%%\citet{RL-bit-by-bit}
%%\citet{guesnerie}
%%\citet{binmore}
%%\citet{grimm-etal-value-equivalence}
%%\citet{littman-sutton-singh}
%%\citet{castillo-reis}
%%\citet{andre-etal-narratives}
%%\citet{barberis-etal-XCAPM}
%%\citet{nagel-xu}
%%\citet{delao-myers}
%%\citet{kase-melosi-rottner}
%%\citet{kozlowski-veldkamp-venky}
%%\citet{ilut-schneider}
%%\citet{giustinelli-manski-molinari}
%%\citet{manski-molinari}
%%\citet{kurz}
%%\citet{iovino-sergeyev}
%%\citet{krusell-smith-rule-of-thumb}
%%\citet{primiceri}
%%\citet{angeletos-huo}
%%\citet{bohren-hauser}
%%\citet{bouchaud-farmer}
%%\citet{glaeser-nathanson}
%%\citet{blume-easley}
%%\citet{colon-bouchaud}
%%\citet{bouchaud-moran-etal}
%%\citet{bouchaud-inelastic}
%%\citet{bouchaud-dessertaine-etal}
%%\citet{bouchaud-knicker-etal}
%%\citet{bouchaud-sharma-etal}
%%\citet{pangallo-etal}
%%\citet{hansen-lunde-nason}
%%\citet{poledna-miess-hommes-rabitsch}
%%\citet{naumannwoleske}
%%\citet{tobin}
%%\citet{mcleay-radia-thomas}
%%\citet{holm-paul-tischbirek}
%%\citet{gatti-etal}

%PLEASE ADD TO REFERENCES IN \url{challenge_bib.bib}

%Papers to cite on belief heterogeneity (WAIT FOR R1)
%\begin{itemize}
%%\item \citet{angeletos_2009} \url{references_request/articles/Heterogeneous_Believes_Models/Angeletos_2009.pdf}
%%\item \citet{angeletos_lao_2013} \url{references_request/articles/Heterogeneous_Believes_Models/Angeletos_Lao_2013.pdf}
%\item \citet{hellwig_2005} \url{references_request/articles/Heterogeneous_Believes_Models/Hellwig_2005.pdf}
%%\item \citet{lao_2010} \url{references_request/articles/Heterogeneous_Believes_Models/Lao_2010.pdf}
%%\item \citet{woodford_2001} \url{references_request/articles/Heterogeneous_Believes_Models/Woodford_2001.pdf}
%\end{itemize}

\small
\setstretch{1.0}{\bibliographystyle{aer}
\bibliography{challenge_bib}
}

\end{document}